\newif\ifconfver
\confverfalse      
\confvertrue        

\newif\ifcutshort      
\cutshorttrue

\newif\ifcutshortlvltwo  
\cutshortlvltwofalse

\ifconfver
    \documentclass[10pt,twocolumn,twoside]{IEEEtran}
    \usepackage{amsfonts}
\else
    \documentclass[11pt,draftcls,onecolumn]{IEEEtran}
\fi

\usepackage{cite}
\usepackage{graphicx}
\usepackage{psfrag}
\usepackage{subfigure}
\usepackage{url}
\usepackage{stfloats}
\usepackage{amsfonts,amssymb,amsmath,bm,paralist,theorem,cite,ifthen,color}
\usepackage{array}
\usepackage{caption}
\usepackage{calc}
\usepackage{upgreek}
\usepackage{subfig}
\usepackage{longtable}
\usepackage{stfloats}
\usepackage{graphicx}
\usepackage{algorithm,algpseudocode}
\usepackage{color}

\newtheorem{lemma}{Lemma}

\theorembodyfont{\rmfamily}

{\begin{list}{}{
    \settowidth{\labelwidth}{\mbox{\textnormal{#1}}}%
    \setlength{\leftmargin}{\labelwidth+\labelsep}}}%
{\end{list}}

\definecolor{orange}{RGB}{255,107,0}

\begin{document}

\bibliographystyle{IEEEtran}

\title{Optimal Real-time Spectrum Sharing between Cooperative Relay and Ad-hoc Networks}


\ifconfver \else {\linespread{1.1} \rm \fi

\author{
{Yin Sun$^\dag$,~\IEEEmembership{Student Member,~IEEE,} Xiaofeng Zhong$^\dag$,~\IEEEmembership{Member,~IEEE,} Tsung-Hui Chang$^\ddag$,~\IEEEmembership{Member,~IEEE,}
Shidong Zhou$^\dag$,~\IEEEmembership{Member,~IEEE,} Jing Wang$^\dag$,~\IEEEmembership{Member,~IEEE,} and Chong-Yung Chi$^\ddag$, ~\IEEEmembership{Senior Member,~IEEE.}}

\thanks{Copyright (c) 2011 IEEE. Personal use of this material is permitted. However, permission to use this material for any other purposes must be obtained from the IEEE by sending a request to pubs-permissions@ieee.org.}
\thanks{
This work was supported by National Basic Research Program of China
(2012CB316002), National S\&T Major Project (2010ZX03005-001-02), China's 863 Project (2009AA011501), National Natural
Science Foundation of China (60832008), Tsinghua Research Funding (2010THZ02-3), and National Science Council, Taiwan, under grant NSC-99-2221-E-007-052-MY3. Yin Sun's work has been supported in part through NSF grant CNS-1012700, and from the Army Research Office MURI W911NF-08-1-0238.
The material in this paper was presented in part in IEEE ICC 2010 and IEEE ICC 2011.}
\thanks{$^\dag$Yin Sun, Xiaofeng Zhong, Shidong Zhou, and Jing Wang were with the State Key Laboratory
on Microwave and Digital Communications, Tsinghua National
Laboratory for Information Science and Technology, and Department of
Electronic Engineering, Tsinghua University, Beijing 100084, China. Yin Sun is now with the Department of Electrical and Computer Engineering, the Ohio State University, Columbus, Ohio 43210, USA.
E-mail: sunyin02@gmail.com,
\{zhongxf,zhousd,wangj\}@tsinghua.edu.cn.}

\thanks{$^\ddag$Tsung-Hui Chang and Chong-Yung Chi are with Institute of Communications Engineering, and Department of Electrical Engineering, National Tsing Hua University,
Hsinchu, Taiwan 30013. E-mail: tsunghui.chang@gmail.com,
cychi@ee.nthu.edu.tw.}
 }
\maketitle

\begin{abstract}\vspace{0cm}
Optimization based spectrum sharing strategies have been widely studied. However, these strategies usually require a great amount of real-time computation and significant signaling delay, and thus are hard to be fulfilled in practical scenarios. This paper investigates optimal real-time spectrum sharing between a cooperative relay network (CRN) and a nearby ad-hoc network. Specifically, we optimize the spectrum access and resource allocation strategies of the CRN so that the average
traffic collision time between the two networks can
be minimized while maintaining a required throughput for the
CRN. The development is first for a frame-level setting, and then is extended to an ergodic setting. For the latter setting, we propose an appealing optimal real-time spectrum sharing strategy via Lagrangian dual optimization. The proposed method only involves a small amount of real-time computation and negligible control delay, and thus is suitable for practical implementations.
Simulation results are presented to demonstrate the efficiency of the proposed strategies.
\end{abstract}

\begin{IEEEkeywords}
Cognitive radio, relay network,
ad-hoc network, resource allocation, real-time control, spectrum
sharing, collision prediction.
\end{IEEEkeywords}

\ifconfver \else \IEEEpeerreviewmaketitle} \fi

\ifconfver \else
\newpage
\fi


\vspace{-0.0cm}
\section{Introduction}\label{sec:intro}
In recent years, spectrum sharing between heterogeneous wireless
networks has been recognized as a crucial technology for improving spectrum efficiency
\cite{Zhao_survey,Servey_spectrum_utilization09} and network
capacity
\cite{Chandrasekhar_Femto_survey08,Hybrid_Cellular_AdHoc_Ton10}.
There are two major models for spectrum sharing presently, namely,
the \emph{open sharing model} and the \emph{hierarchical access
model} \cite{Zhao_survey,Servey_spectrum_utilization09}. In the open sharing model, each network
has equal right to access the same spectrum band, e.g., the unlicensed
band, and there is no strict constraint on the interference level
from one network to its neighbors. In the hierarchical access model
which consists of a primary network and a secondary network, the
secondary network, i.e., cognitive radio, dynamically accesses the
spectrum provided that the primary
users' transmission is almost not affected \cite{Zhao_survey}. In
either model, the inter-network
interference make spectrum sharing a challenging task, especially
when there is no explicit coordination between the coexisting
networks.

To address this interference issue, cognitive spectrum access strategies have been proposed \cite{Zhao_Jsac07,
Xi_ZhangJsac08,Geirhofer_ComMag,Geirhofer_Jsac08,Zhaoqianchuan_TSP08,GeirhoferExperiment09,
ZhaoTong_Jsac11,LifengLai_MAC11} for the hierarchical access model. While these works focus on MAC-layer spectrum access,
there have been works focusing on physical-layer resource allocation of
secondary networks, where strict constraints are imposed to limit
the induced interference to the primary users; see
\cite{Peng_Wang07,RuiZhangJSTSP08,PowerOFDMCogTSP09} and also
\cite{RuiZhangMag10} for an overview. Joint optimization of
spectrum access and resource allocation was studied in
\cite{Geirhofer_Mobile_computing} for an open sharing model that
considers spectrum sharing between an uplink orthogonal frequency-division
multiplexing (OFDM) system and an ad-hoc
network.

However, some implementation issues were rarely considered
in these optimization based spectrum sharing strategies. First, real-time optimization can be computationally quite demanding for realistic wireless networks \cite{cellular_adhoc_tradeoffjsac09}. Second, spectrum sensing and channel estimation are usually performed
at spatially separate nodes, which requires information exchange between these nodes before solving the optimization
problem. The resultant computation and signaling procedure usually lead to significant control delay, making these strategies hardly be fulfilled in practical scenarios. Therefore, spectrum sharing strategies with little real-time calculation and small signaling delay are of great importance for practical applications.

\begin{figure*}[!t] \centering
    \resizebox{0.75\textwidth}{!}{
    \includegraphics{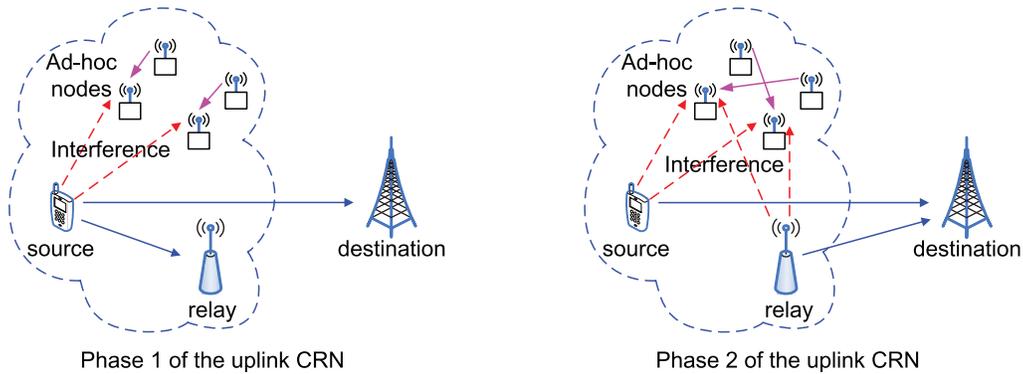}}
    \caption{System model. In Phase 1 (left plot), the source node broadcasts information to the relay and destination, and the transmitted signal
     interferes with the nearby ad-hoc links;
     in Phase 2 (right plot), the source and relay transmit signals to the destination simultaneously and both transmitted signals interfere with the ad-hoc links.}
    \label{fig1}
    \vspace{0cm}
\end{figure*}

In this paper, we study spectrum sharing between an uplink broadband cooperative relay
network (CRN) and an ad-hoc network, as illustrated in Fig.
\ref{fig1}.
The CRN is composed of
a source node (e.g., a mobile terminal), a relay node, and a distant
destination node (e.g., a base station (BS)).
The CRN adopts a two-phase transmission protocol for
each frame: in the first phase, the source broadcasts
an information message to the relay and BS; in the second phase, the
relay employs a broadband decode-and-forward (DF) strategy \cite{Madsen05,Liang07, Vandendorpe08,Vandendorpe09} to forward the
message. The source will transmit a new message to the BS in the
second phase as well. In order to communicate with the
distant BS, the source and relay transmit signals with peak powers,
which, however, will induce strong interference to nearby ad-hoc
links operating over the same spectrum band. The ad-hoc
transmitters, e.g., wireless sensor nodes, have relatively low
transmission powers due to their short communication ranges, and
thus their interference to the relay and destination can be treated
as noise. Such an asymmetrical interference scenario is known as the
``near-far effect" \cite{Chandrasekhar_Femto_survey08}.
Our contributions in
this paper are summarized as follows:
    \vspace{0.3cm}
\begin{enumerate}
\item We first consider the case that the source and relay nodes perform spectrum sensing at the start of each frame. By modeling the
ad-hoc traffic in each ad hoc band as independent binary continuous-time Markov chains
(CTMC) \cite{BK:Resnick}, the average traffic collision time between the two networks can be obtained based on collision prediction.
We formulate a frame-level spectrum sharing design problem that jointly optimizes physical-layer resource allocation and MAC-layer cognitive spectrum access of the CRN such that the
average traffic collision time is minimized, while maintaining a required throughput for the
CRN.

\item The formulated spectrum sharing problem is a difficult nonconvex
optimization problem with no closed-form expression for the
objective function. We first derive the optimal spectrum access
strategy, based on which, we show that the resource allocation problem
can be reformulated as a convex optimization problem. To solve this
problem in a low-complexity manner, we present a Lagrangian dual optimization method, which has a linear complexity with respect to the number of
sub-channels.

\item As common issues of existing frame-level transmission control strategies, the developed frame-level spectrum sharing strategy requires excessive computation and signaling process, which may cause considerable control delay and is not suitable for real-time implementations. To overcome these issues, we further formulate an \emph{ergodic} spectrum sharing design problem based on a long term average CRN achievable
rate and a long term average traffic collision time. By exploiting the Lagrangian dual optimization solution, we develop a low-complexity, real-time implementation method for obtaining the optimal spectrum sharing strategy. The proposed strategy is appealing because most computation tasks are accomplished off-line, leaving only simple tasks for real-time computations.
In addition, the computation and signaling procedure are carefully designed to maximally reduce the control delay. This method can accommodate an additional spectrum sensing in Phase 2 of each frame to further improve the accuracy of collision prediction.
\end{enumerate}

%

The spectrum sharing strategies proposed in this paper differs from that reported in \cite{Geirhofer_Mobile_computing} in three aspects: First, our interference metric is more practical in the considered strong interference scenarios (see Remark 1 in Section \ref{sec3A} for more details). Second, the
spectrum sharing design problem of our DF based CRN is more difficult compared to that of point-to-point uplink system considered in \cite{Geirhofer_Mobile_computing}. Finally, an optimal real-time implementation method is proposed for the ergodic spectrum sharing design strategy, which is never reported in the literature before.

The proposed strategies may provide potential spectrum sharing solutions for various application scenarios; e.g., the coexistence between unlicensed
WiMAX (IEEE 802.16) and WiFi (IEEE 802.11)
\cite{unlicensed_wimax07}, the coexistence between a relay-assisted
cellular network and ad-hoc networks including mobile ad-hoc
networks \cite{cellular_adhoc_tradeoffjsac09} and peer-to-peer
communication networks \cite{peer_to_peer07}, and, in the domain of
military communications, the coexistence between the broadband
tactical backbone network and local ad-hoc networks, such as sensor
networks and tactical mobile ad-hoc networks \cite{Wimax_milcom10}.

For ease of later use, let us define the following notations: The
probability of event $A$ is denoted by $\Pr\{A\}$, and the
probability of event $A$ conditioned on event $B$ is denoted by
$\Pr\{A|B\}$. The indicator function of event $A$ is given by $\textbf{1}(A)$.  $\mathbb{E}_\omega\{X\}$ represents expectation of $X$ over random variable
$\omega$, and $\mathbb{E}\{X|Z=z\}$ is the conditional expectation of $X$
given $Z = z$. We denote $\parallel\bm x\parallel_2$ as the Euclidean norm of vector $\bm x$, and denote $\pi(S)$ as the size (measure) of a set,
e.g., $\pi([a,b])=b-a$. The projections of $x$ on the sets $[0,\infty)$ and $[0,y]$ are denoted by $[x]^+ = \max\{x,0\}$ and $[x]_0^y = \min\{\max\{x,0\},y\}$, respectively.

The remaining parts of this paper are organized as follows: In
Section \ref{sec:prob statement}, the system description and the
formulation of frame-level spectrum sharing problem
are presented. Section \ref{sec3} presents a Lagrangian dual optimization
method to resolve the frame-level spectrum sharing problem. The ergodic spectrum sharing problem
and its real-time implementation method are presented in Section
\ref{sec4}. Conclusions are drawn in Section \ref{sec14}.

\vspace{0.0cm}
\section{System Description and Problem Formulation}\label{sec:prob statement}\vspace{0.2cm}


\subsection{Signal Model}
\begin{figure*}[!t] \centering
    \resizebox{0.75\textwidth}{!}{
    \includegraphics{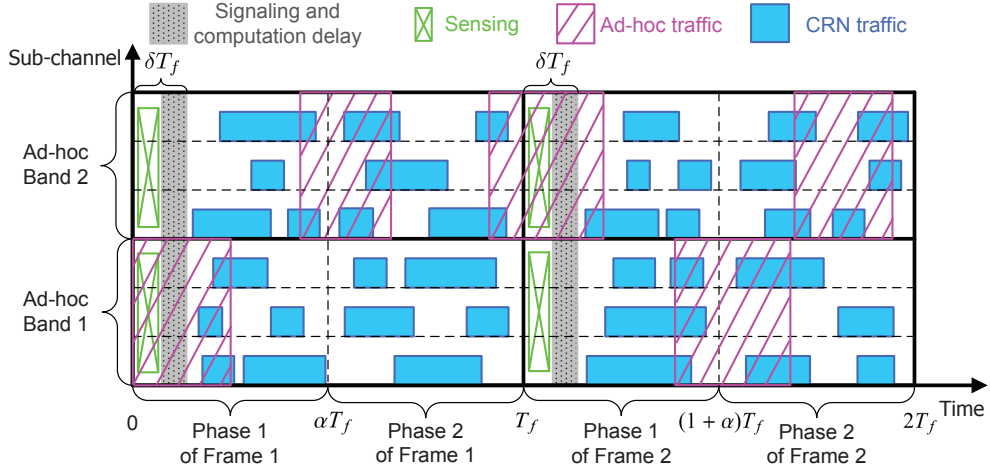}}
    \caption{Time-frequency transmission structure of the CRN and the ad-hoc network.
    In this figure, the number of CRN sub-channels is 6 ($N=6$) and the number of ad-hoc bands is 2 ($M=2$).}
    \label{fig5}
    \vspace{-0cm}
\end{figure*}
We assume that the CRN employs a broadband multi-carrier air
interface, where all nodes transmit and receive signals over $N$
parallel sub-channels, denoted by $\mathcal {N}=\{1,2,\ldots,N\}$. The term ``sub-channel" here represents either a frequency
subband or a group of consecutive subcarriers in OFDM systems \cite{Zhao_survey}. The ad-hoc links operate in $M$ non-overlapped frequency bands,
denoted by $\mathcal {M}=\{1,\ldots,M\}$. Moreover, the $m$th
ad-hoc band overlaps with the sub-channels of the CRN in the set
$\mathcal {N}_m\subseteq \mathcal {N}$ for $m=1,\ldots,M$, where
$\mathcal {N}_m$ satisfies $\bigcup_{m = 1}^M\mathcal {N}_m=\mathcal
{N}$ and $\mathcal {N}_p\bigcap\mathcal {N}_q = {\O}$ for $p\neq q$.
An example with $M=2$ and $N=6$ is illustrated in Fig. \ref{fig5}.

In the temporal
domain, all nodes in the CRN transmit and receive signals in a
frame-by-frame manner, where each frame has a fixed duration $T_f$.
The source node performs spectrum sensing at the start of each frame to detect the ACTIVE/IDLE state of each ad-hoc band. Perfect sensing and negligible sensing duration is assumed in this paper, which is reasonable for moderate sensing
signal-to-noise ratio (SNR) and common value of the frame duration $T_f$ \footnote{This assumption was confirmed in \cite{Geirhofer_Jsac08} for a WLAN energy detector. When the sensing SNR is 5 dB, a sensing duration of
less than 5$\mu$s is sufficient to achieve a detection error probability of $10^{-5}$. If we choose a frame duration from 500$\mu$s to 2ms,
the sensing duration is fairly small.}.
Since the computation capability of the source node is quite limited,
the source node cannot determine the optimal spectrum sharing design. Therefore, it forwards the obtained sensing results to the
destination. The destination node computes the optimal spectrum access and
resource allocation strategies of the CRN, and feeds them back to the
source and relay nodes. As one may have noticed, the computation and signaling procedure may cause considerable control delay between spectrum sensing and data transmission. In this work, we will explicitly take into account this control delay in the spectrum sharing design. As illustrated in Fig. 2, we use $\delta T_f$ to denote this control delay, where $\delta\in[0,1)$. For large value of $\delta$, the transmission time of the CRN  $(1-\delta) T_f$ becomes quite small, leading to a significant performance degradation. In Section IV, a real-time spectrum sharing strategy with negligible control delay will be presented.

In practice, the relay node operates in a half-duplex mode
\cite{Madsen05}. Therefore, each frame consists of 2 phases: In
Phase 1, the source transmits signal to the relay and destination
via a broadcast channel; in Phase 2, the source transmits a new
information message, and, at the same time, the relay uses the DF
relay strategy to forward the information message received in
Phase 1 to the destination, which forms a multiple-access channel.
These scenarios are illustrated in Fig. \ref{fig1}. The time
durations of Phase 1 and Phase 2 are set to $\alpha T_f$ and
$(1-\alpha)T_f$, respectively, where $\alpha\in(\delta,1)$.

We assume that the source and relay nodes can switch on and off
their transmissions \emph{freely} over each sub-channel. Let
$\mathbb{I}_n^{(1)}\subseteq[\delta T_f,\alpha T_f]$ denote the time set that the source node transmits over the $n$th sub-channel in Phase 1, and
$\mathbb{I}_n^{(2)}\subseteq[\alpha T_f,T_f]$ denote the time set that the
source and relay nodes transmit over the $n$th sub-channel in Phase 2, for $n=1,\ldots,N$. As the example in Fig. \ref{fig5} shows, $\mathbb{I}_n^{(1)}$ and $\mathbb{I}_n^{(2)}$ each may be \emph{a union of several disjoint transmission time intervals}. Note that the source node cannot transmit during $[0,\delta T_f]$, owing to aforementioned control delay. For convenience, let us
define
\begin{eqnarray}\label{eq2}
\theta_n^{(1)}=\frac{\pi(\mathbb{I}_n^{(1)})}{T_f},~~~\theta_n^{(2)}=\frac{\pi(\mathbb{I}_n^{(2)})}{T_f},
\end{eqnarray}
which represent the fractions of the CRN transmission time in Phase
1 and 2 of each frame, respectively.

We assume that the wireless channels of source-relay,
source-destination, and relay-destination links are block-fading
\cite{networkinformation10}, which means that the channel
coefficients remain static within each frame, and can change from
one frame to another. Let $h_n^{i,j}$ denote the frequency response
of sub-channel $n$ between transmitter $i$ and receiver $j$, where
$i\in\{s,r\}$ and $j\in\{r,d\}$ ($i\neq j$), in which $s,$ $r,$ $d$ stand
for the source node, relay node and destination node, respectively. The
interference plus noise at the relay and destination nodes are modeled as
independent, zero mean, circularly symmetric complex Gaussian random
variables, with $N_n^r$ and $N_n^d$ denoting the respective peak power
spectral densities (PSD) over the $n$th sub-channel (i.e., the weak interference from the ad-hoc network to the CRN is treated as noise). Hence, the quality of
the wireless links can be characterized by the normalized channel
power gains $g_n^{s,r}\triangleq {|h_n^{s,r}|^2}/{N_n^rW}$,
$g_n^{s,d}\triangleq { |h_n^{s,d}|^2}/{N_n^dW}$, and
$g_n^{r,d}\triangleq {|h_n^{r,d}|^2}/{N_n^dW}$, where $W$ is the
bandwidth of each sub-channel.
For broadband DF CRN with $N$ parallel
sub-channels, the following rate is achievable \cite[Eq. (45)]{Madsen05}
\begin{align}\label{eq84}
\!\!\!\!\!\!\!\!\!\!\!\!&R_{CRN}=W\min\left\{\sum_{n=1}^N\left[{\theta_n^{(1)}}
\log_2\left(1+\!\frac{
P_{s,n}^{(1)}\max\{g_n^{s,r},g_n^{s,d}\}}{\theta_n^{(1)}}\!\right)\right.\right.\nonumber\\&
~~~~~~~~~~~~~~~~~~~~~~~~~~~~~~+\left.{\theta_n^{(2)}} \log_2\left(1+\frac{
P_{s,n}^{(2)}g_n^{s,d}}{\theta_n^{(2)}}\right)\!\right],
\nonumber\\&~~~~~~~~~~~~~~~~~~~\sum_{n=1}^N\left[
{\theta_n^{(1)}} \log_2\left(1+\frac{
P_{s,n}^{(1)}g_n^{s,d}}{\theta_n^{(1)}}\right)\right.\nonumber\\&~~~~~~~~~~~~~~~\left.\left.+ \theta_n^{(2)}
 \log_2\left(1+\frac{P_{s,n}^{(2)}g_n^{s,d}+
P_{r,n} g_n^{r,d}}{\theta_n^{(2)}}\right)\right]\right\},\!\!\!\!\!
\end{align}
where $P_{s,n}^{(1)}$ and $P_{s,n}^{(2)}$ denote the transmission powers of the source over sub-channel $n$ in Phase 1 and Phase 2, respectively; and $P_{r,n}$ is the transmission power of the relay over sub-channel $n$ in Phase 2.
The achievable rate $R_{CRN}$ in (\ref{eq84}) is a concave function of the transmission power and time variables
$\{P_{s,n}^{(1)},P_{s,n}^{(2)},P_{r,n},\theta_n^{(1)},$ $\theta_n^{(2)},n\in\mathcal {N}\}$ \cite[p. 89]{BK:BoydV04}.

The ad-hoc traffic over the $m$th band is modeled as an independent,
stationary binary CTMC $X_m(t)$, where $X_m(t)=1$ ($X_m(t)=0$)
represents an ACTIVE (IDLE) state at time $t$. The holding periods of both ACTIVE and IDLE states are exponentially
distributed with rate parameters $\lambda$ and $\mu$,
respectively. The probability transition matrix of the CTMC model of
band $m$ is given by \cite[p. 391]{BK:Resnick}
\begin{align} \label{eq15}
P(t)=\frac{1}{\lambda+\mu}\left[\begin{array}{l l}\mu+\lambda
e^{-(\lambda+\mu)t}&\lambda-\lambda
e^{-(\lambda+\mu)t}\\\mu-\mu e^{-(\lambda+\mu)t}&\lambda+\mu
e^{-(\lambda+\mu)t}\end{array}\right],
\end{align} where the element in the $i$th row and $j$th column of $P(t)$ stands for
the transition probability $\Pr\{X_m(t+\tau)=j-1|X_m(\tau)=i-1\}$
for $i, j\in\{1,2\}$. This CTMC model has been used in many theoretical
spectrum sharing studies and verified by hardware
tests; see
\cite{Geirhofer_ComMag,Geirhofer_Jsac08,Zhaoqianchuan_TSP08,ZhaoTong_Jsac11,Geirhofer_Mobile_computing,GeirhoferExperiment09}.
In practice, the parameters $\lambda$ and $\mu$ can be estimated by
monitoring the ad-hoc traffic in idle frames of the CRN
\cite{Geirhofer_ComMag}.

\subsection{Traffic Collision Prediction and Interference Metric}
We utilize the average traffic collision time between the CRN and the ad-hoc network as the metric of interference experienced by the ad-hoc links. Since the ad-hoc nodes are near the source and relay nodes, the ad-hoc links would
suffer from communication errors, whenever the ad-hoc transmission happens to
collide with the CRN traffic \footnote{In general, the transmission error probability of an ad-hoc band is an increasing function of the average traffic collision time. In particular, if the ad-hoc transmission is uncoded in the physical layer or the ad-hoc
code block length is quite short, the transmission error probability caused by traffic collisions is approximately proportional to the average traffic collision time.}. Let $x_m\in\{0,1\}$
denote the sensing outcome for the $m$th ad-hoc band, i.e.,
$X_m(0)=x_m$. Given $x_m$, one can predict the average traffic
collision time based on the CTMC model in \eqref{eq15}.
Specifically, the average traffic collision time over
the $m$th ad-hoc band is given by
\begin{eqnarray}\label{eq21}
&&\mathbb{E}\left\{\int_{\bigcup\limits_{n\in\mathcal
{N}_m}\mathbb{I}_n^{(1)}} \textbf{1}(X_m(t)=1) dt\right.\nonumber\\
&&~~\left.\left.+\int_{\bigcup\limits_{n\in\mathcal
{N}_m}\mathbb{I}_n^{(2)}} \textbf{1}(X_m(t)=1) dt\right| X_m(0)=x_m\right\},
\end{eqnarray}
where $\textbf{1}(X_m(t)=1)$ is the indicator function of event $X_m(t)=1$.
It is worthwhile to note that $\bigcup_{n\in\mathcal {N}_m}\mathbb{I}_n^{(i)}$ in \eqref{eq21} is the time set that the CRN is transmitting in at least one sub-channel in $\mathcal {N}_m$, where $\mathcal {N}_m$ is the set of sub-channels overlapping with the $m$th ad-hoc band. This reflects the fact that, in the considered strong interference scenario, the ad-hoc transmission in the $m$th band is disrupted, even if the CRN transmits in only one sub-channel of $\mathcal {N}_m$.

Each of the expectation terms in \eqref{eq21}
can be calculated as follows
\begin{eqnarray}\label{eq27}
&&\mathbb{E}\left.\left\{\int_{\bigcup\limits_{n\in\mathcal
{N}_m}\mathbb{I}_n^{(i)}} \textbf{1}\left(X_m(t)=1\right) dt\right| X_m(0)=x_m\right\}\nonumber\\
\!\!\!\!\!\!\!\!\!&&=\int_{\bigcup\limits_{n\in\mathcal
{N}_m}\mathbb{I}_n^{(i)}} \mathbb{E}\left\{X_m(t)=1 | X_m(0)=x_m\right\}dt\nonumber\\
&&=\int_{\bigcup\limits_{n\in\mathcal {N}_m}\mathbb{I}_n^{(i)}}
\Pr\left\{X_m(t)=1|X_m(0)=x_m\right\} dt.
\end{eqnarray}
By \eqref{eq27}, the total traffic collision time summed over all the
ad-hoc bands is given by
\begin{eqnarray}\label{eq105}
\!\!\!\!\!\!\!\!\!\!\!\!\!&&I
=\sum_{m=1}^M\left[\!\int_{\bigcup\limits_{n\in\mathcal
{N}_m}\mathbb{I}_n^{(1)}}  \!\!\!\!\!\!
\Pr\!\left\{X_m(t)\!=\!1|X_m(0)\!=\!x_m\right\} dt
\right.\nonumber\\&&~~~\left.+
\int_{\bigcup\limits_{n\in\mathcal {N}_m}\mathbb{I}_n^{(2)}}
\!\!\!\!\Pr\left\{X_m(t)=1|X_m(0)=x_m\right\}
dt\right].
\end{eqnarray}

\subsection{Frame-level Spectrum Sharing Design}

The goal of the CRN is to optimize the source and relay's
transmission powers $P_{s,n}^{(1)},P_{s,n}^{(2)}$ and $P_{r,n}$, and
their spectrum access strategies $\mathbb{I}_n^{(1)}$ and
$\mathbb{I}_n^{(2)}$, such that the total traffic collision time
in \eqref{eq105} is minimized, while a minimum uplink throughput
$R_{\min}$ can be maintained. This joint spectrum access and
resource allocation design problem can be formulated as the following
optimization problem:
\begin{center}
\fbox{\parbox[]{0.97 \linewidth}{
\begin{subequations} \label{PFC_subproblem}
\begin{align}
\!\!\!\!(\sf P)\min_{\substack{P_{s,n}^{(1)},P_{s,n}^{(2)},P_{r,n},~\\
\mathbb{I}_n^{(1)},\mathbb{I}_n^{(2)}}}
\!\!\!\!&\frac{I}{T_f}\\
{\rm s.t.}~~~~&R_{CRN}\geq R_{\min},\label{eq111}\\
\!\!\!\!& \sum_{n = 1}^N \left[P_{s,n}^{(1)}+ P_{s,n}^{(2)}\right]\leq P^s_{\max},\label{eq100}\\
\!\!\!\!&\sum_{n = 1}^N P_{r,n}\leq P^r_{\max},\label{eq101}\\
\!\!\!\!& P_{s,n}^{(1)},P_{s,n}^{(2)},P_{r,n}\geq 0,~~n=1,\ldots,N,\label{eq4}\\
\!\!\!\!& \mathbb{I}_n^{(1)}\subseteq[\delta T_f,\alpha T_f],~~\mathbb{I}_n^{(2)}\subseteq[\alpha T_f,T_f],\nonumber\\
\!\!\!\!&~~~~~~~~~~~~~~~~~~~~~~~~~~~n=1,\ldots,N,\label{eq44}\\
\!\!\!\!&\pi(\mathbb{I}_n^{(1)}) = \theta_n^{(1)}T_f,~~\pi(\mathbb{I}_n^{(2)})=\theta_n^{(2)}T_f,\nonumber\\
\!\!\!\!&~~~~~~~~~~~~~~~~~~~~~~~~~~~n=1,\ldots,N, \label{eq64}
\end{align}
\end{subequations}
}}
\end{center}
where $R_{CRN}$ and $I$ are given by \eqref{eq84} and \eqref{eq105},
respectively, \eqref{eq64} follows from \eqref{eq2},
$P^s_{\max}$ and $P^r_{\max}$ in \eqref{eq100} and \eqref{eq101} denote the power
constraints at the source and relay nodes, respectively.

\section{Optimal Frame-level Spectrum Sharing Solution}\label{sec3}

Problem $(\sf P)$ is difficult to solve because the objective function $I$ has no closed-form expression. Fortunately, this issue can be resolved by analyzing the optimal transmission time sets in the two phases of each frame, i.e., $I_n^{(1)}$ and $I_n^{(2)}$, from which Problem $(\sf P)$ can be reformulated as a convex optimization problem, as
we will present in this section. A Lagrangian dual
optimization method is also proposed to obtain an
optimal solution of $(\sf P)$ efficiently.

\subsection{Reformulation of Problem $(\sf P)$}\label{sec3A}

The key idea that makes this convex reformulation possible is to
examine the optimal spectrum access $\mathbb{I}_n^{(1)}$ and
$\mathbb{I}_n^{(2)}$ in Problem $(\sf P)$. In particular, it can be
shown (in Lemma \ref{lem1} below) that the optimal spectrum access must satisfy the following two principles:
\begin{enumerate}
\item In both Phase 1 and Phase 2, the source and relay nodes should transmit as soon (late)
as possible if the sensing outcome is IDLE (ACTIVE);

\item The CRN should have identical spectrum access strategy for
the sub-channels overlapping with the same ad-hoc band; that is,
$\mathbb{I}_p^{(i)}=\mathbb{I}_q^{(i)}$ for all $p,q\in\mathcal
{N}_m$, where $m\in\mathcal {M}$ and $i\in\{1,2\}$.
\end{enumerate}
Principle 2) shares the same flavor of interference alignment technique
in \cite{Jafar_MIMO_X} since both of them align the transmissions to reduce the interference to the ad-hoc links.
Let us define
\begin{equation}\label{eq74}
\hat\theta_m^{(i)}\triangleq\max\left\{\theta_n^{(i)},n\in \mathcal
{N}_m\right\}
\end{equation}
as the largest transmission time fraction over the sub-channels
$\mathcal {N}_m$ in phase $i$. The optimal spectrum access strategy is given by the following lemma, which is proved in
Appendix \ref{Transmission_time_placement}.

\begin{figure*}[!t]
    \centering
        \resizebox{0.75\textwidth}{!}{\includegraphics{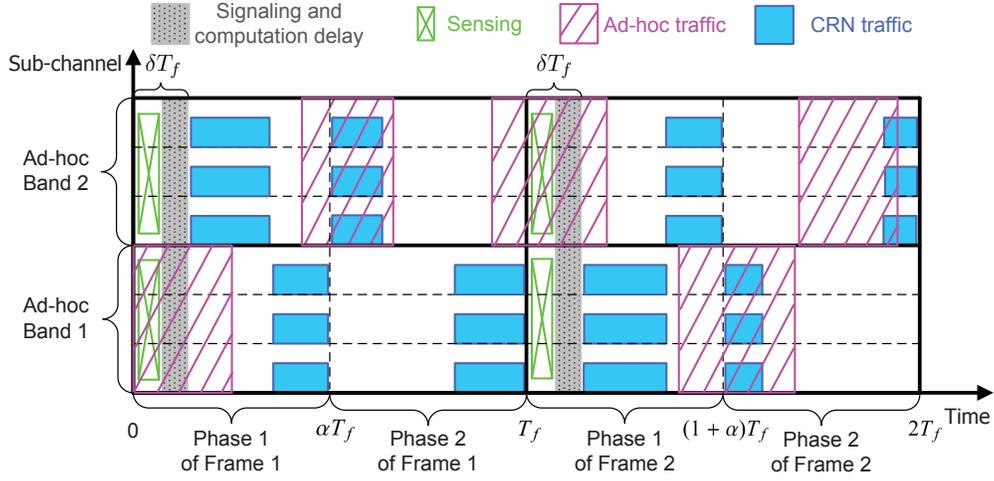}}
        \caption{An example illustrating the spectrum access strategy of Lemma
        \ref{lem1}.
        The sensing outcomes of Frame 1 are $x_1=1$ and $x_2=0$, and the sensing outcomes of Frame 2 are $x_1=0$ and $x_2=1$.}
        \label{fig2}
        \vspace{-0.5cm}
\end{figure*}

\begin{lemma}\label{lem1}
For any given transmission time fractions
$\{\hat\theta_m^{(1)}\in[0,\alpha-\delta],\hat\theta_m^{(2)}\in[0,1\!-\!\alpha]\}_{m=1}^M$ and transmission power
$\{P_{s,n}^{(1)},P_{s,n}^{(2)},P_{r,n}\}_{n=1}^N$
in Problem $(\sf P)$, we have that:
\begin{enumerate}
\item The
optimal spectrum access strategy in Phase 1 is given by
$\mathbb{I}_n^{(1)}=[\delta T_f,(\delta+\hat\theta_m^{(1)})T_f]$ for all $n\in\mathcal
{N}_m$ if the sensing outcome is $x_m = 0$, and is given by
$\mathbb{I}_n^{(1)}=[(\alpha-\hat\theta_m^{(1)})T_f, \alpha T_f]$
for all $n\in\mathcal {N}_m$ if the sensing outcome is $x_m = 1$;

\item The
optimal spectrum access strategy in Phase 2 is given by
$\mathbb{I}_n^{(2)}=[\alpha T_f,(\alpha+\hat\theta_m^{(2)} )T_f]$
for all $n\in\mathcal {N}_m$ if the sensing outcome is $x_m = 0$,
and is given by $\mathbb{I}_n^{(2)}=[(1-\hat\theta_m^{(2)})T_f,
T_f]$ for all $n\in\mathcal {N}_m$ if the sensing outcome is $x_m = 1$.
\end{enumerate}
\end{lemma}
An example that illustrates the spectrum access strategy of Lemma
\ref{lem1} is shown in Fig. \ref{fig2}.

\emph{Remark 1:} A similar result of Lemma \ref{lem1} was reported in \cite{Geirhofer_Mobile_computing} for a different interference metric, which cumulates the traffic collisions for all the sub-channels overlapping with the same ad-hoc band. For example,
$\left.\sum_{n\in\mathcal
{N}_m}\mathbb{E}\left\{\int_{\mathbb{I}_n^{(i)}} X_m(t) dt\right|
X_m(0)=x_m\right\}$ represents the
average traffic collision time over the $m$th ad-hoc band in \cite{Geirhofer_Mobile_computing} instead of our interference metric given in \eqref{eq27}. Our interference metric is
more practical than the one reported in \cite{Geirhofer_Mobile_computing}, because, in strong interference scenario, the ad-hoc transmission in the $m$th band is disrupted, no matter that the CRN is transmitting in either one or more sub-channels of $\mathcal {N}_m$. Since transmitting in more sub-channels of $\mathcal {N}_m$ will not further worsen the interference, the CRN should transmit over all the sub-channels of $\mathcal {N}_m$ simultaneously to increase data rate, which is different from the result of \cite{Geirhofer_Mobile_computing}.

%

According to Lemma
\ref{lem1}, the integration region in each term of \eqref{eq105} is a simple time interval.
In order to simplify \eqref{eq105}, we define for $\theta \in [0,\alpha-\delta]$,
\begin{eqnarray}\label{eq16}
\!\!\!\!\!\!\!\!\!\!\!\!\!\!\!\!\!\!\!&&~~~\phi_{(1)}
(\theta;x_m=0)\!\!\!\!\!\!\!\!\!\!\!\nonumber\\
\!\!\!\!\!\!\!\!\!\!\!\!\!\!\!\!\!\!\!&&=
\int_{[\delta T_f,(\delta+\theta) T_f]}\Pr (X_m(t)=1|X_m(0)=0)d t\nonumber\\
\!\!\!\!\!\!\!\!\!\!\!\!\!\!\!\!\!\!\!&&=\frac{\lambda}{\lambda\!+\!\mu}
\left\{\theta\!+\!\frac{1}{(\lambda\!+\!\mu)T_f}e^{\!-({\lambda\!+\!\mu})\delta
T_f}
\left[e^{-{(\lambda+\mu})\theta T_f}\!-\!1\!\right]\!\right\}\!T_f\!,
\\
\!\!\!\!\!\!\!\!\!\!\!\!\!\!\!\!\!\!\!&&~~~\phi_{(1)}(\theta;x_m=1)\!\!\!\!\!\!\!\!\!\!\!\nonumber\\
\!\!\!\!\!\!\!\!\!\!\!\!\!\!\!\!\!\!\!&&=
\int_{[(\alpha-\theta) T_f,\alpha T_f]}\Pr (X_m(t)=1|X_m(0)=1)d t\nonumber\\
\!\!\!\!\!\!\!\!\!\!\!\!\!\!\!\!\!\!\!&&=\frac{\lambda}{\lambda\!+\!\mu}\!\left\{\!\theta
\!+\!\frac{\mu}{\lambda}\frac{1}{(\lambda\!+\!\mu)
T_f}e^{\!-({\lambda\!+\!\mu})\alpha
T_f}\!\!\left[\!e^{({\lambda+\mu})\theta T_f}\!-\!1\!\right]\!\!\right\}\!T_f,
\label{eq166}
\end{eqnarray}
and define for $\theta \in [0, 1-\alpha]$,
\begin{eqnarray}
\!\!\!\!\!\!\!\!\!\!\!\!\!\!\!\!\!\!\!&&~~~\phi_{(2)}(\theta;x_m=0)\!\!\!\!\!\!\!\!\!\!\!\nonumber\\
\!\!\!\!\!\!\!\!\!\!\!\!\!\!\!\!\!\!\!&&=\!\int_{[\alpha
T_f,(\theta+\alpha) T_f]} \Pr(X_m(t)\!=\!1|X_m(0)=0)d
t\!\nonumber\\
\!\!\!\!\!\!\!\!\!\!\!\!\!\!\!\!\!\!\!&&=\!\frac{\lambda}{\lambda+\mu}\!\left\{\!\theta +\frac{1}{(\lambda\!+\!\mu)T_f}e^{\!-({\lambda\!+\!\mu})\alpha
T_f}\!\!\left[\!e^{-({\lambda\!+\!\mu})\theta T_f}-1\!\right]\!\!\right\}\!T_f,
\end{eqnarray}
\begin{eqnarray}
\!\!\!\!\!\!\!\!\!\!\!\!\!\!\!\!\!\!\!&&~~~\phi_{(2)}(\theta;x_m=1)\!\!\!\!\!\!\!\!\!\!\!\nonumber\\
\!\!\!\!\!\!\!\!\!\!\!\!\!\!\!\!\!\!\!&&=\int_{[T_f-\theta
T_f,T_f]} \Pr(X_m(t)=1|X_m(0)=1)d
t\!\nonumber\\
\!\!\!\!\!\!\!\!\!\!\!\!\!\!\!\!\!\!\!&&=\!\frac{\lambda}{\lambda\!+\!\mu}\!\left\{\!\theta +\!\frac{\mu}{\lambda}\frac{1}{(\lambda\!+\!\mu)T_f}e^{-({\lambda\!+\!\mu})T_f}
\!\left[\!e^{({\lambda+\mu})\theta T_f}\!-\!1\right]\!\!\right\}\!T_f.\label{eq17}
\end{eqnarray}
The interference metric in
\eqref{eq105} can be simplified as
\begin{eqnarray} \label{eq6}
I = \sum_{m=1}^M \left[ \phi_{(1)}
\left(\hat\theta_m^{(1)};x_m\right)+\phi_{(2)}
\left(\hat\theta_m^{(2)};x_m\right)\right].
\end{eqnarray}
It is easy to verify that the functions $\phi_{(1)}(\theta;x)$ and
$\phi_{(2)}(\theta;x)$ in \eqref{eq16}-\eqref{eq17} are strictly
increasing and strictly convex functions of $\theta$. Thus $I$ in
\eqref{eq6} is a convex function of $\hat\theta_m^{(1)}$ and
$\hat\theta_m^{(2)}$. Further, the control delay $\delta T_f$
degrades the interference mitigation performance of the spectrum access strategy in Lemma \ref{lem1}, as $\phi_{(1)}(\theta;0)$ is strictly increasing in $\delta$.

On the other hand, it follows from
\eqref{eq84}, \eqref{eq74} and Lemma \ref{lem1} that the constraint
\eqref{eq111} of $(\sf P)$ can be equivalently expressed as
\begin{eqnarray}\label{eq43}
R_1\geq R_{\min},~R_2\geq R_{\min},
\end{eqnarray}
where
\begin{eqnarray}\label{eq7}
R_{1}\!=\!\!\!\!\!\!\!\!\!\!\!&&
W\!\sum_{m\in\mathcal {M}} \sum_{n\in\mathcal {N}_m}
\left[{\hat\theta_m^{(1)} }\log_2\!\left(1+\!\frac{ P_{s,n}^{(1)}
\max\{g_n^{s,r},g_n^{s,d}\}}{\hat\theta_m^{(1)} }\!\right)\right.\nonumber\\
&&~~~~\left.+{\hat\theta_m^{(2)}} \log_2\left(1+\frac{
P_{s,n}^{(2)} g_n^{s,d}}{\hat\theta_m^{(2)} }\right)\right],\\
R_{2}=\!\!\!\!\!\!\!\!\!\!\!&&
W\sum_{m\in\mathcal {M}}\sum_{n\in\mathcal {N}_m}\left[
{\hat\theta_m^{(1)}}\log_2\left(1+\frac{ P_{s,n}^{(1)}
g_n^{s,d}}{\hat\theta_m^{(1)} }\right)\right. \nonumber\\
&&~~~~\left.+\hat\theta_m^{(2)}\log_2\left(1+\frac{P_{s,n}^{(2)} g_n^{s,d}+
P_{r,n}g_n^{r,d}}{\hat\theta_m^{(2)} }\right)\right].\label{eq8}
\end{eqnarray}
By \eqref{eq6} and \eqref{eq43}, Problem $(\sf P)$ is equivalent to the following problem:
\begin{subequations} \label{PFC_subproblem1}
\begin{align}
\!\!\!\!\min_{\substack{P_{s,n}^{(1)},P_{s,n}^{(2)},P_{r,n},\\
\hat\theta_m^{(1)},\hat\theta_m^{(2)}}} &\frac{1}{T_f}\sum_{m=1}^M \!\!\left[ \phi_{(1)}
\!\left(\hat\theta_m^{(1)};x_m\right)\!+\!\phi_{(2)}
\!\left(\hat\theta_m^{(2)};x_m\right)\right]\!\!\label{eq11}\\
{\rm s.t.}~~~~~&R_1\geq R_{\min} \label{eq20}\\
\!\!\!\!&R_2\geq R_{\min} \label{eq54}\\
\!\!\!\!& \sum_{n = 1}^N \left[P_{s,n}^{(1)}+ P_{s,n}^{(2)}\right]\leq P^s_{\max}\label{eq3} \\
\!\!\!\!& \sum_{n = 1}^N P_{r,n}\leq P^r_{\max} \label{eq56}\\
\!\!\!\!& P_{s,n}^{(1)},P_{s,n}^{(2)},P_{r,n}\geq 0,~~n=1,\ldots,N\label{eq4}\\
\!\!\!\!& 0\leq \hat\theta_m^{(1)}\leq \alpha-\delta,~0\leq
\hat\theta_m^{(2)}\leq 1-\alpha,\nonumber\\
\!\!\!\!&~~~~~~~~~~~~~~~~~~~~~~~~~~ m=1,\ldots, M,\label{eq57}
\end{align}
\end{subequations}
which is a convex optimization problem. While problem
\eqref{PFC_subproblem1} can be solved by interior-point methods, we
present in the next subsection a low-complexity Lagrangian dual
optimization method.

\subsection{Lagrangian Dual Optimization Method for Problem \eqref{PFC_subproblem1}}\label{sec3B}

Suppose that problem \eqref{PFC_subproblem1} is strictly feasible.
Then, according to the Slater's condition \cite{BK:BoydV04}, the
strong duality holds for \eqref{PFC_subproblem1}. Hence we can
alternatively consider the following dual optimization problem
\begin{eqnarray} \label{eq32}
\max_{\zeta,\sigma,\varepsilon,\eta\geq0}\left\{\min_{(P_{s,n}^{(1)}
,P_{s,n}^{(2)} , P_{r,n}, \hat\theta_m^{(1)}, \hat\theta_m^{(2)})\in
\mathcal {V}} L\right\},
\end{eqnarray}
where $\mathcal {V}\!\triangleq\!\{\!\left.(P_{s,n}^{(1)}
,P_{s,n}^{(2)} , P_{r,n}, \hat\theta_m^{(1)},
\hat\theta_m^{(2)})\right|0\!\leq\!
\hat\theta_m^{(1)}\!\leq\!\alpha\!-\!\delta,$ $0\!\leq\!
\hat\theta_m^{(2)}\!\leq\! 1\!-\!\alpha, P_{s,n}^{(1)}
,P_{s,n}^{(2)} , P_{r,n} \!\geq\! 0, n\!\in\!\mathcal
{N},m\!\in\!\mathcal {M}\}$,
\begin{align} \label{eq51}
\!\!\!\!L &=\frac{1}{T_f}\sum_{m=1}^M \left[ \phi_{(1)}
\left(\hat\theta_m^{(1)};x_m\right)+\phi_{(2)}
\left(\hat\theta_m^{(2)};x_m\right)\right]\nonumber\\
\!\!\!\!&+
\frac{\zeta}{W}({R}_{\min}\!-\!R_1)\!+\!\frac{\sigma}{W}({R}_{\min}\!-\!R_2)\nonumber\\
\!\!\!\!&+\varepsilon \left[\sum_{n=1}^N
(P_{s,n}^{(1)}\!+\!P_{s,n}^{(2)})\!-\!P_{\max}^s \right]\!+\!\eta\!
\left[\sum_{n=1}^N P_{r,n}\!-\!P_{\max}^r \right]\!,\!\!
\end{align}
is the partial Lagrangian \cite{Chiang07ProcIEEE} of \eqref{PFC_subproblem1},
$\zeta\!\geq0$, $\sigma\geq0,$ $\varepsilon\geq0,$ and $\eta\geq0$ are the dual
variables associated with the constraints \eqref{eq20},
\eqref{eq54}, \eqref{eq3}, and \eqref{eq56}, respectively. As will be
shown, the inner minimization problem of \eqref{eq32} has
closed-form solutions for $P_{s,n}^{(1)}$, $P_{s,n}^{(2)}$ and
$P_{r,n}$ for $n=1,\ldots,N$, and $\hat\theta_m^{(1)}$ and
$\hat\theta_m^{(2)}$ for $m=1,\ldots,M$. Hence, the computational
complexity for solving the inner problem is only linear with respect
to $N$ and $M$. Moreover, the outer maximization problem of
\eqref{eq32} only involves four optimization variables
($\zeta,\sigma,\varepsilon,\eta$), which is much smaller than the
number of variables of the primal problem \eqref{PFC_subproblem1}.

Suppose that a dual variable $\bm\nu\triangleq(\zeta,\sigma,\varepsilon,\eta)^T$ is
given. Let us present the closed-form solutions of the inner
minimization problem of \eqref{eq32}. Because the inner problem is
convex, the optimal $(P_{s,n}^{(1)} ,P_{s,n}^{(2)} , P_{r,n},
\hat\theta_m^{(1)},$ $\hat\theta_m^{(2)})$ for fixed dual variable $\bm\nu$ must satisfy the Karush-Kuhn-Tucker (KKT) conditions \cite{BK:BoydV04} of the inner problem, which can be expressed as:
{\small
\begin{eqnarray} \label{eq55}
\frac{\partial
L}{\partial  P_{s,n}^{(1)}}&&\!\!\!\!\!\!\!\!\!\!\!\!=-\frac{{\zeta
\max\{g_n^{s,r},g_n^{s,d}\}}}{\left(1\!+\max\!\{g_n^{s,r}\!,g_n^{s,d}\}\frac{P_{s,n}^{(1)}\!}{\hat\theta_m^{(1)}}\right)\ln2}-\frac{{\sigma
g_n^{s,d}}}{\left(1\!+\!g_n^{s,d}\frac{P_{s,n}^{(1)}}{\hat\theta_m^{(1)}}\right)\ln2}\nonumber\\
&&~~~~~~~~~~~~~~~~~~~~~~~~~~~~~~~~+\varepsilon\!\left\{\!\!\!\begin{array}{l}
\!=\! 0~ \textrm{if}~P_{s,n}^{(1)}\!\! >\!0 \\  \!\geq\! 0~  \textrm{if}~P_{s,n}^{(1)}\!\!=\! 0\end{array}\right.\!\!\!\!\!\!\!\!\!\!\!\!\!\\
\frac{\partial
L}{\partial  P_{s,n}^{(2)}}&&\!\!\!\!\!\!\!\!\!\!\!\!=-\frac{{\zeta
g_n^{s,d}}}{\left(1\!+\!g_n^{s,d}\frac{P_{s,n}^{(2)}}{\hat\theta_m^{(2)}}\right)\ln2}-\frac{{\sigma
g_n^{s,d}}}{\left(1\!+\!g_n^{s,d}\frac{P_{s,n}^{(2)}
}{\hat\theta_m^{(2)}}\!+\!g_n^{r,d}\frac{P_{r,n}}{\hat\theta_m^{(2)}}\right)\ln2}\nonumber\\
&&~~~~~~~~~~~~~~~~~~~~~~~~~~~~~~~~+\varepsilon\!
\left\{\!\!\!\begin{array}{l} = 0~ \textrm{if}~P_{s,n}^{(2)}\!>\!0\\
\geq 0~  \textrm{if}~P_{s,n}^{(2)}\!= \!0\end{array}\right.\label{eq66}\!\!\!\!\!\!\!\!\!\!\!\!\!\\
\frac{\partial
L}{\partial  P_{r,n}}&&\!\!\!\!\!\!\!\!\!\!\!\!=-\frac{{\sigma
g_n^{r,d}}}{\left(1+g_n^{s,d}\frac{P_{s,n}^{(2)}
}{\hat\theta_m^{(2)}}+g_n^{r,d}\frac{P_{r,n}
}{\hat\theta_m^{(2)}}\right)\ln2}\nonumber\\
&&~~~~~~~~~~~~~~~~~~~~~~~~~~~~~~~~+\eta\left\{\!\!\!\begin{array}{l}=0~ \textrm{if}~P_{r,n}>0
\\\geq0~ \textrm{if}~P_{r,n}=0  \end{array}\right.\label{eq70}\!\!\!\!\!\!\!\!\!\!\!\!\!\\
\frac{\partial
L}{\partial \hat\theta_m^{(1)}}&&\!\!\!\!\!\!\!\!\!\!\!\! =-\zeta\!\!\sum_{n\in
\mathcal
{N}_m}\!\!\!f\!\!\left(\max\{g_n^{s,r},g_n^{s,d}\}\frac{P_{s,n}^{(1)}}{\hat\theta_m^{(1)}}\right)\!-\!\sigma\!\!\sum_{n\in
\mathcal
{N}_m}\!\!\!f\!\!\left(g_n^{s,d}\frac{P_{s,n}^{(1)}}{\hat\theta_m^{(1)}}\right)\nonumber\\
&&~~~~~~~\!+\!\frac{\partial\phi_{(1)}
\left( \hat\theta_m^{(1)};
x_m\right)}{\partial\hat\theta_m^{(1)}}\!\left\{\!\!\!\begin{array}{l}\leq0~
\textrm{if}~\hat\theta_m^{(1)}=\alpha\!-\!\delta\\=0~
\textrm{if}~\hat\theta_m^{(1)}\!\!\in\!(0,\alpha\!-\!\delta)
 \\\geq0~ \textrm{if}~\hat\theta_m^{(1)}=0\end{array}\right.~\label{eq62}\\
\frac{\partial
L}{\partial \hat\theta_m^{(2)}}&&\!\!\!\!\!\!\!\!\!\!\!\!=-\zeta\!\!\sum_{n\in \mathcal
{N}_m}\!\!\!f\!\!\left(\!g_n^{s,d}\frac{P_{s,n}^{(2)}}{\hat\theta_m^{(2)}}\!\right)\!-\!\sigma\!\!\sum_{n\in
\mathcal {N}_m}\!\!\!f\!\!\left(g_n^{s,d}\frac{P_{s,n}^{(2)}
}{\hat\theta_m^{(2)}}\!+\!g_n^{r,d}\frac{P_{r,n}}{\hat\theta_m^{(2)}}\!\right)\nonumber\\
&&~~~~~~~\!+\!\frac{\partial\phi_{(2)}\left(
\hat\theta_m^{(2)};x_m\right)}{\partial\hat\theta_m^{(2)}}\!\left\{\!\!\!\begin{array}{l}\leq0~
\textrm{if}~\hat\theta_m^{(2)}=1\!-\!\alpha\\=0~
\textrm{if}~\hat\theta_m^{(2)}\!\in\!(0,1\!-\!\alpha) \\\geq0~
\textrm{if}~\hat\theta_m^{(2)}=0\end{array}\right.\label{eq63}
\end{eqnarray}$\!\!$}
where $f(x)\triangleq\log_2\left(1+x\right)-\frac{x}{(1+x)\ln2}$ in \eqref{eq62} and \eqref{eq63}.

We first solve (\ref{eq55}) to obtain the optimal ratio
$\frac{P_{s,n}^{(1)}}{\hat\theta_m^{(1)}}$. Specifically, if
$P_{s,n}^{(1)}>0$, then equality in (\ref{eq55}) holds, and an optimal $\frac{P_{s,n}^{(1)}}{\hat\theta_m^{(1)}}$
is equal to the positive root $x$ of the following quadratic
equation
\begin{eqnarray} \label{eq59}
\frac{\zeta
\max\{g_n^{s,r},g_n^{s,d}\}}{1+\max\{g_n^{s,r},g_n^{s,d}\}x}+\frac{\sigma
g_n^{s,d}}{1+g_n^{s,d}x}=\varepsilon\ln 2.
\end{eqnarray}
If \eqref{eq55} has no positive root, then $\frac{P_{s,n}^{(1)}}{\hat\theta_m^{(1)}}=0$. Next, let us find the
optimal $\frac{P_{s,n}^{(2)}}{\hat\theta_m^{(2)}}$ and
$\frac{P_{r,n}}{\hat\theta_m^{(2)}}$ by solving \eqref{eq66} and
\eqref{eq70}. If $P_{r,n}>0$, the equality in \eqref{eq70} holds, and the optimal $\frac{P_{s,n}^{(2)}}{\hat\theta_m^{(2)}}$
and $\frac{P_{r,n}}{\hat\theta_m^{(2)}}$ can be obtained from
\eqref{eq66} and \eqref{eq70} as
\begin{eqnarray}\label{eq67}
&&\frac{P_{s,n}^{(2)}}{\hat\theta_m^{(2)}}=\left[\frac{\zeta}{(\varepsilon-\eta
g_n^{s,d}/g_n^{r,d})\ln2}-\frac{1}{g_n^{s,d}}\right]^+,\\
&&\frac{P_{r,n}}{\hat\theta_m^{(2)}}=\frac{\sigma}{\eta\ln2
}-\frac{1}{g_n^{r,d}}-\frac{g_n^{s,d}}{g_n^{r,d}}\frac{P_{s,n}^{(2)}}{\hat\theta_m^{(2)}},\label{eq68}
\end{eqnarray}
where $[x]^+ = \max\{x,0\}.$ Instead, if $P_{r,n}=0$, we obtain from \eqref{eq66} and \eqref{eq70} that
\begin{eqnarray}\label{eq76}
&&\frac{P_{s,n}^{(2)}}{\hat\theta_m^{(2)}}=\left[\frac{\zeta+\sigma}{\varepsilon\ln2}-\frac{1}{g_n^{s,d}}\right]^+,\\
&&\frac{P_{r,n}}{\hat\theta_m^{(2)}}=0,\label{eq75}
\end{eqnarray}
where \eqref{eq76} is actually the water-filling solution when the source directly communicates with the destination without the use of relay.

The optimal $\hat\theta_m^{(1)}$ and $\hat\theta_m^{(2)}$ can be
obtained by solving \eqref{eq62} and \eqref{eq63}, respectively,
provided that the optimal
$\frac{P_{s,n}^{(1)}}{\hat\theta_m^{(1)}}$,
$\frac{P_{s,n}^{(2)}}{\hat\theta_m^{(2)}}$ and
$\frac{P_{r,n}}{\hat\theta_m^{(2)}}$ have been obtained from
\eqref{eq59}-\eqref{eq75}. By substituting
$\frac{P_{s,n}^{(1)}}{\hat\theta_m^{(1)}}$,
$\frac{P_{s,n}^{(2)}}{\hat\theta_m^{(2)}}$ and
$\frac{P_{r,n}}{\hat\theta_m^{(2)}}$ into \eqref{eq62}-\eqref{eq63},
and by the definitions of  $\phi_{(1)}(\theta;x)$ and
$\phi_{(2)}(\theta;x)$ in \eqref{eq16}-\eqref{eq17}, we can obtain
the optimal values of $\hat\theta_m^{(1)}$ and $\hat\theta_m^{(2)}$
as follows: For $x_m = 0$, we have \footnotemark[1]\footnotetext[1]{
For the notational simplicity in \eqref{eq12}-\eqref{eq13}, we have
extended the definition of the natural logarithm $\ln(x)$ to that
with $\ln(x)=-\infty$ for $x\in(-\infty,0]$ .}
\begin{eqnarray} \label{eq12}
\!\!\!\!\!\!\!\!\!\!&&\hat\theta_m^{(1)}=
\left[\!-\frac{1}{(\lambda\!+\!\mu)T_f}\!\ln\!\left\{1\!-\!\frac{\lambda\!+\!\mu}{\lambda}\!\sum_{n\in
\mathcal {N}_m}\!\!\left[\!{\sigma}
f\!\!\left(\!g_n^{s,d}\frac{P_{s,n}^{(1)}}{\hat\theta_m^{(1)}}\!\right)\!\!\!\!\right.\right.\right.\nonumber\\
\!\!\!\!\!\!\!\!\!\!&&~~~~\left.\left.\left.+{\zeta}
f\!\!\left(\max\{g_n^{s,r},g_n^{s,d}\}\frac{P_{s,n}^{(1)}}{\hat\theta_m^{(1)}}\!\right)\right]\right\}-\delta\right]_0^{\alpha-\delta},
\\
\!\!\!\!\!\!\!\!\!\!&&\hat\theta_m^{(2)}=
\left[-\frac{1}{(\lambda\!+\!\mu)T_f}\!\ln\!\left\{1\!-\!\frac{\lambda\!+\!\mu}{\lambda}\!
\!\sum_{n\in \mathcal
{N}_m}\!\!\left[\!\zeta f\!\!\left(\!g_n^{s,d}\frac{P_{s,n}^{(2)}}{\hat\theta_m^{(2)}}\!\right)\!\!\!\!\!\right.\right.\right.\nonumber\\
\!\!\!\!\!\!\!\!\!\!&&~~~~\left.\left.\left.+{\sigma}
f\left(g_n^{s,d}\frac{P_{s,n}^{(2)}}{\hat\theta_m^{(2)}}+g_n^{r,d}\frac{P_{r,n}}{\hat\theta_m^{(2)}}\!\right)\right]
\right\}-\alpha\right]_0^{1-\alpha},\label{eq37}
\end{eqnarray}
\begin{algorithm}
  \caption{The proposed Lagrangian dual optimization algorithm for solving $(\sf P)$}\label{alg1}
\begin{algorithmic}[1]
  \State \textbf{Input} system parameters $(N,M,\alpha,P^s_{\max},P^r_{\max}, T_f,$ $R_{\min})$, the ad-hoc traffic parameters
  $\lambda,\mu$, the channel quality $\{g_n^{s,r},g_n^{s,d},g_n^{r,d}\}_{n=1}^N$, the sensing outcome
  $\{x_m\}_{m=1}^M$, the computation and signaling delay parameter $\delta$, and a solution accuracy $\epsilon$.
  \State Set the iteration number $k=1$; initialize the dual variable $\bm\nu_1$.
  \State Compute the optimal $\{P_{s,n}^{(1)},P_{s,n}^{(2)},P_{r,n}\}_{n=1}^N$ and $\{\hat\theta_m^{(1)},\hat\theta_m^{(2)}\}_{m=1}^M$ according to \eqref{eq59}-\eqref{eq13}.
  \State Update the dual variable $\bm\nu_{k+1}$ according to \eqref{eq58} and \eqref{eq99}.
  \State If $\parallel\bm\nu_{k+1}-\bm\nu_{k}\parallel_2\leq\epsilon$, go to Step 6; otherwise, set $k=k+1$ and return to Step 3.
  \State \textbf{Output} the optimal primal solution $\{P_{s,n}^{(1)},P_{s,n}^{(2)},P_{r,n}\}_{n=1}^N$ and
  $\{\hat\theta_m^{(1)},\hat\theta_m^{(2)}\}_{m=1}^M$.
  The optimal spectrum access strategy $\{\mathbb{I}_n^{(1)},\mathbb{I}_n^{(2)}\}_{n=1}^{N}$ can be obtained by Lemma \ref{lem1}.
\end{algorithmic}
\end{algorithm}
where $[x]_0^y=\min\{\max\{x,0\},y\}$, and for $x_m = 1$, we have
\begin{eqnarray}\label{eq38}
\!\!\!\!\!\!\!\!\!\!&&\hat\theta_m^{(1)}=\left[\!\alpha\!+\!\frac{1}{(\lambda\!+\!\mu)T_f}\!\ln\!\left\{\!\frac{\lambda\!+\!\mu}{\mu}\!
\!\sum_{n\in \mathcal {N}_m}\!\left[\!{\sigma}
f\!\!\left(\!g_n^{s,d}\frac{P_{s,n}^{(1)}}{\hat\theta_m^{(1)}}\!\right)\!\!\right.\right.\right.\nonumber\\
\!\!\!\!\!\!\!\!\!\!&&~~~~~~\left.\left.\left.+{\zeta}f\!\!\left(\max\{g_n^{s,r},g_n^{s,d}\}\frac{P_{s,n}^{(1)}}
{\hat\theta_m^{(1)}}\!\right)\!\right]\!-\!\frac{\lambda}{\mu}\!\right\}\!\right]_0^{\alpha-\delta},\\
\!\!\!\!\!\!\!\!\!\!&&\hat\theta_m^{(2)}=
\left[1+\frac{1}{(\lambda\!+\!\mu)T_f}\!\ln\!\left\{\!\frac{\lambda\!+\!\mu}{\mu}
\!\sum_{n\in \mathcal
{N}_m}\!\left[\!\zeta f\!\left(g_n^{s,d}\frac{P_{s,n}^{(2)}}{\hat\theta_m^{(2)}}\right)\right.\right.\right.\!\!\!\!\nonumber\\
\!\!\!\!\!\!\!\!\!\!&&~~~\left.\left.\left.+{\sigma}f\left(g_n^{s,d}\frac{P_{s,n}^{(2)}}{\hat\theta_m^{(2)}}
+g_n^{r,d}\frac{P_{r,n}}{\hat\theta_m^{(2)}}\!\right)\right]\!
-\!\frac{\lambda}{\mu}\!\right\}\!\right]_0^{1-\alpha}.\!\!\!\label{eq13}
\end{eqnarray}
By substituting (\ref{eq12})-(\ref{eq13})
into (\ref{eq59})-(\ref{eq75}), the optimal
$P_{s,n}^{(1)},P_{s,n}^{(2)}, P_{r,n}$ can then be obtained.

What remains for solving \eqref{eq32} is to optimize the dual variable $\bm\nu=(\zeta,\sigma,\varepsilon,\eta)^T$ for the outer maximization problem. In view of that the dual function of \eqref{PFC_subproblem1} (the optimal value of the inner problem of \eqref{eq32}) may not be differentiable \cite{BK:Bertsekas}, we consider to update $\bm\nu$ using the subgradient method \cite{Subgradient_Boyd}.
Specifically, at the $k$th iteration, the subgradient method updates $\bm\nu$ by \cite{Subgradient_Boyd}
\begin{eqnarray}\label{eq58}
\bm\nu_{k+1} =\left[\bm\nu_{k} + s_k \bm h(\bm\nu_{k})\right]^+,
\end{eqnarray}
where the subscript $k$ denotes the iteration number, $s_k$ is the
step size of the $k$th iteration, and $\bm h(\bm \nu_{k})$ is the
subgradient of the dual function,
which is given by \cite{BK:Bertsekas}
\begin{eqnarray}\label{eq99}
\bm h(\bm \nu_{k})=\left[\begin{array}{l}(R_{\min}-R_{1}^\star)/W\\(R_{\min}-R_{2}^\star)/W\\
\sum_{n=1}^N
\left(P_{s,n}^{(1)\star}+P_{s,n}^{(2)\star}\right)-P_{\max}^s\\\sum_{n=1}^N
P_{r,n}^\star -P_{\max}^r\end{array}\right],
\end{eqnarray}
where $P_{s,n}^{(1)\star}$, $P_{s,n}^{(2)\star}$ and $P_{r,n}^\star$
are the optimal solution of the inner minimization problem
\eqref{eq32} at iteration $k$, and $R_1^\star$ and $R_2^\star$ are
the corresponding rate values in \eqref{eq7} and \eqref{eq8},
respectively. It has been shown that the subgradient updates in
\eqref{eq58} converge to the optimal dual point $\bm\nu^\star$ as
$k\rightarrow\infty$, provided that the step size $s_k$ is chosen
according to a diminishing step size rule \cite{Subgradient_Boyd}.
The convergence speed of the subgradient method can be improved if
one further considers the acceleration techniques in
\cite{BK:Bertsekas,BK:Bazaraa06,BK:Shor85}. In Algorithm \ref{alg1},
we summarize the proposed Lagrangian dual optimization algorithm of
$(\sf P)$.

\begin{table} \caption{Parameter setting for Fig. \ref{fig4}.} \label{tab2} \centering
\begin{tabular}{|c|c|c|c|c|c|c|c|c|}
\hline \!Parameter\! & $N$ & $M$ & $\lambda T_f$ & $\mu T_f$ & $\!P^s_{\max}\!$ & $\!P^r_{\max}\!$ & $\alpha$ & $\delta$\\
\hline Value & 2 & 2  & 1 & 1& 1 & 1 & 0.5 & 0.1\\
\hline\hline
\!Parameter\! & $x_1$ & $x_2$ & $g_1^{s,d}$ & $g_1^{s,r}$ & $g_1^{r,d}$  & $g_2^{s,d}$ & $g_2^{s,r}$ & $g_2^{r,d}$  \\
\hline Value & 0 & 1 & 0.4 & 1.3 & 1.3 & 0.5 & 1.4 & 1.4  \\
\hline
\end{tabular}
\end{table}

\subsection{Simulation Results}\label{sec3C}
In this subsection, we provide some simulation results to examine
the performance of the proposed CRN spectrum sharing strategy in $\sf
(P)$. The parameters used in the simulations are listed in Table
\ref{tab2}. Since $N=M=2$, we simply set $\mathcal{N}_1=\{1\}$
and $\mathcal{N}_2=\{2\}$.

We compare the proposed CRN spectrum sharing strategy
with two degenerated strategies, namely, the \emph{relay-free} strategy
and the \emph{sensing-free} strategy.
Similar to the work in \cite{Geirhofer_Mobile_computing}, the
relay-free strategy only considers the direct uplink transmission from the source to the
destination. To implement this strategy, we simply set
$g_n^{s,r}=g_n^{r,d}=0$ for Problem $(\sf P)$.

\begin{figure*}[!t]
\begin{center}
{\subfigure[][]{\resizebox{0.45\textwidth}{!}{\includegraphics{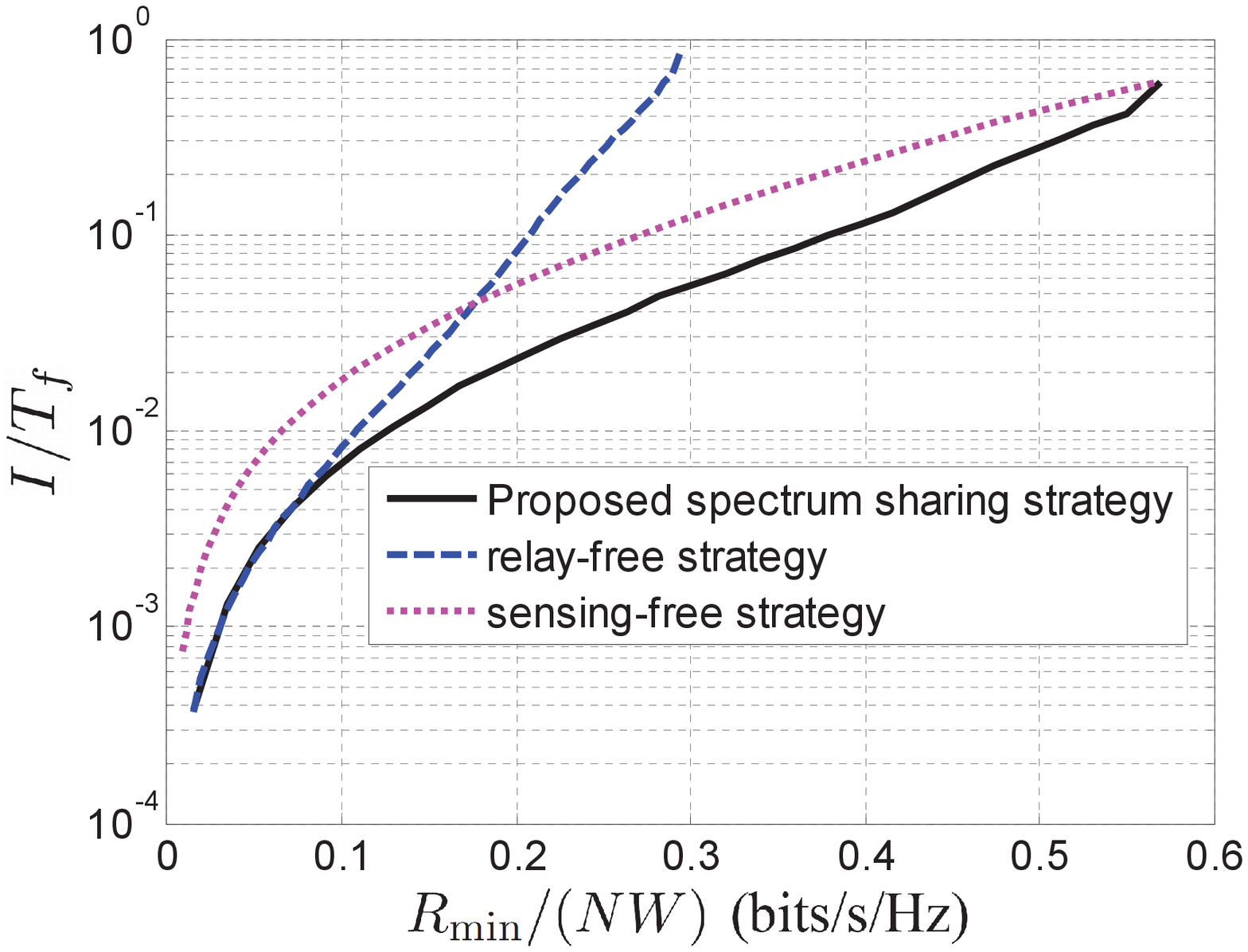}}}}
    \hspace{+0.2cm}
{\subfigure[][]{\resizebox{0.45\textwidth}{!}{\includegraphics{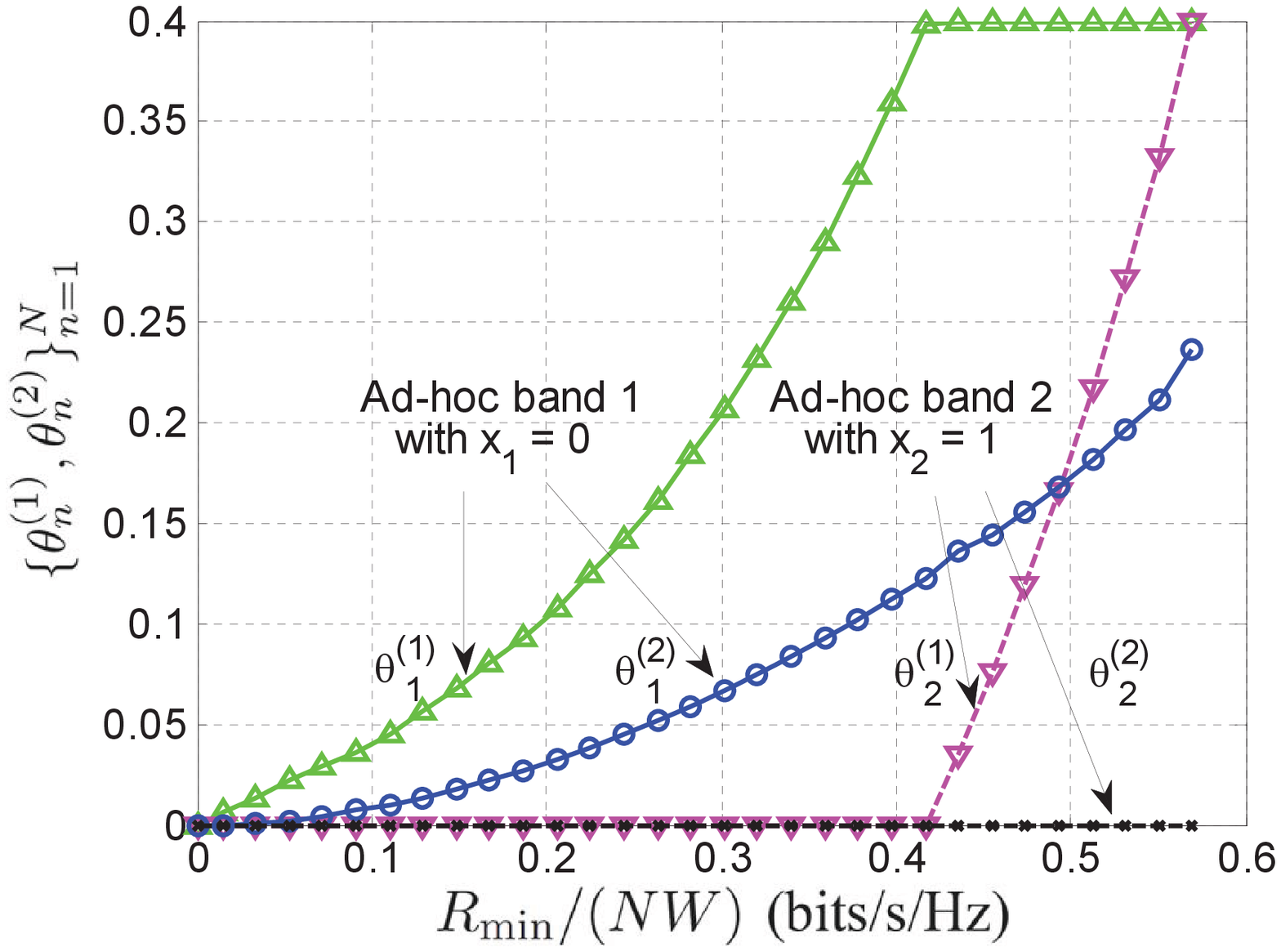}}}
    \vspace{0.5cm}}
\end{center}\vspace{-0.7cm}
\caption{ Simulation results of the proposed frame-level transmission control strategy of the CRN with
parameters given in Table \ref{tab2}. (a) Normalized
average collision time ($I/T_f$) versus required uplink spectrum
efficiency $R_{\rm min}/(NW)$, (b) optimized transmission time
fractions $\{\theta_n^{(1)},\theta_n^{(2)}\}_{n=1}^N$ versus
required uplink spectrum efficiency $R_{\rm min}/(NW)$.}\vspace{-0.5cm}
\label{fig4}
\end{figure*}

In the sensing-free strategy, the average traffic collision time in \eqref{eq105} becomes
\begin{eqnarray}
\hat{I}
\!\!\!\!\!\!\!\!\!\!&&=\sum_{m=1}^M\left[\!\int_{\bigcup_{n\in\mathcal
{N}_m}\mathbb{I}_n^{(1)}}
\Pr\!\left\{X_m(t)\!=\!1\right\} dt\right.\nonumber\\
&&~~\left.+
\int_{\bigcup_{n\in\mathcal {N}_m}\mathbb{I}_n^{(2)}}
\Pr\left\{X_m(t)=1\right\}
dt\right],
\end{eqnarray}
due to the lack of sensing result $X_m(0)=x_m$. It further reduces to
\begin{eqnarray}\label{eq1}
\frac{\hat{I}}{T_f} = \sum_{m=1}^M\frac{\lambda}{\lambda+\mu}\left(\hat{\theta}_m^{(1)}+\hat{\theta}_m^{(2)}\right),
\end{eqnarray}
by means of the interference alignment principle in Lemma \ref{lem1}.
Then, the sensing-free strategy is obtained by solving \eqref{PFC_subproblem1} with the objective function in \eqref{eq11} replaced by \eqref{eq1}.


Figure \ref{fig4}(a) presents the performance comparison results of
normalized average collision time ($I/T_f$) versus required uplink
spectrum efficiency $R_{\rm  min}/(NW)$. We observe from this figure
that for $R_{\rm min}/(NW)$ $<0.07$ bits/s/Hz, the relay-free strategy
exhibits comparable performance with the proposed strategy; whereas
for $R_{\rm min}/(NW)\geq 0.07$ bits/s/Hz, the proposed strategy yields
a smaller average traffic collision time. The performance
improvement is attributed to the DF relay techniques. The sensing-free strategy always generates more traffic
collisions than the proposed strategy, because it does not utilize the
spectrum sensing outcomes. 

Figure \ref{fig4}(b) displays the optimal transmission time
fractions $\{\theta_n^{(1)},\theta_n^{(2)}\}_{n=1}^N$ of $(\sf P)$
versus required uplink spectrum efficiency $R_{\rm min}/(NW)$.
The CRN only transmits over sub-channel 1 when $R_{\min}/(NW)\leq0.42$ bits/s/Hz, because the sensing outcomes are $x_1=0$ and $x_2=1$. For
$R_{\min}/(NW)>0.42$ bits/s/Hz, the CRN starts to transmit over both
sub-channels to achieve more stringent throughput
requirement $R_{\rm min}/(NW)$. The maximal value of $\theta_n^{(1)}$ is $\alpha-\delta = 0.4$, because the computation and signaling delay is $\delta T_f= 0.1T_f$. On the other hand, it is interesting that $\theta_n^{(2)}$ never achieves its maximum $1-\alpha=0.5$, even at the largest feasible value of $R_{\rm min}/(NW)$. This is because the transmission time $\theta_n^{(2)}$ only contributes to $R_2$, but not to $R_1$. When $\theta_1^{(1)}=\theta_2^{(1)}=0.4$, $R_1$ achieves its maximum and $R_{CRN} = \min\{R_1,R_2\}$ is constrained by $R_1$, the CRN cannot increase $R_1$ by increasing $\theta_n^{(2)}$.



\section{Ergodic Spectrum Sharing Design and Real-time Implementation} \label{sec4}
In the previous section, the optimal spectrum sharing design is obtained in each frame. As we mentioned in Section \ref{sec:intro}, such a frame-level spectrum sharing strategy may encounter several implementation issues. First, the CRN has to solve Problem $(\sf P)$ within every frame, which may be computationally too demanding for realistic wireless networks. Moreover, the destination node has to collect the spectrum sensing outcome from the source and/or relay nodes in order to
solve $(\sf P)$. After solving $(\sf P)$, the destination node has
to send the solutions back to the source and relay nodes. The computation and signaling procedure may cause a considerable control delay $\delta T_f$, leaving very short time for data transmission.

These issues intrigue us to investigate more practical spectrum sharing strategies  that allow real-time implementations. The key ideas are 1) to reduce the amount of real-time computations and 2) to decrease the computation and signaling delay. Our approach is to consider an \emph{ergodic} resource allocation problem.
Recall from Section \ref{sec3A} that the frame-level design problem \eqref{PFC_subproblem1} can be solved by the Lagrangian dual optimization method. We will show next that for the ergodic spectrum sharing design problem, one can compute the optimal dual variable $\bm\nu^\star$ off-line, and only some simple tasks are left for real-time computation.


After obtaining $\bm\nu^\star$, we need to compute the transmission parameters based on real-time spectrum sensing and channel estimation results. These real-time computation tasks are fulfilled carefully in order to minimize the computation and signaling delay $\delta T_f$.

Another benefit of this ergodic setting is that it allows one more spectrum sensing at the beginning of Phase 2 to improve the accuracy of collision prediction. This, however, cannot be exploited in the frame-level setting, because Problem $(\sf P)$ must be solved before this additional sensing in Phase 2 is carried out.


\subsection{Ergodic Spectrum Sharing Design Problem}\label{sec41}
\begin{figure*}[!t]
    \centering
        \resizebox{0.75\textwidth}{!}{\includegraphics{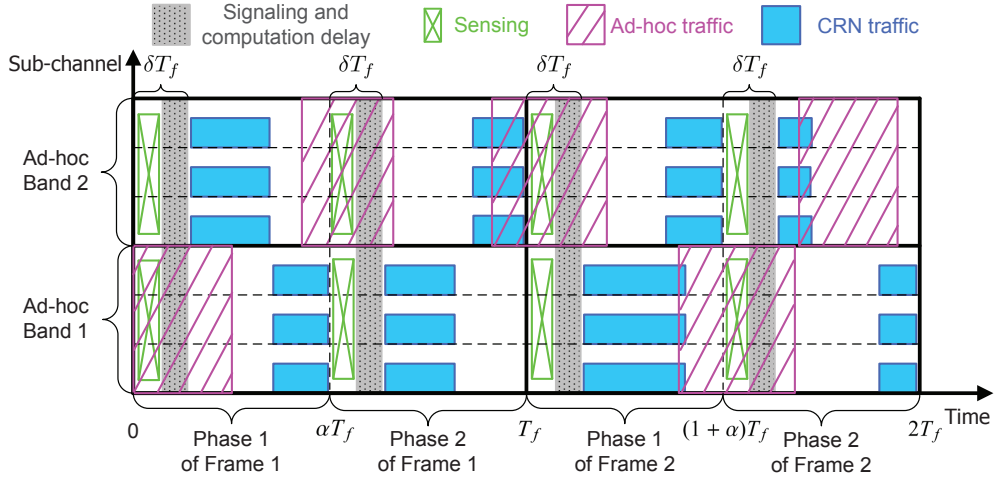}}
        \caption{Time-frequency transmission structure by Lemma \ref{lem3}. Here the sensing outcomes of Frame 1 are $x_1=1$, $x_2=0$, $y_1=0$ and $y_2=1$, and the the sensing outcomes of Frame 2 are $x_1=0$, $x_2=1$, $y_1=1$ and $y_2=0$.}
        \label{fig99}
        \vspace{-0.0cm}
\end{figure*}

Suppose that both the source and relay nodes perform one extra
spectrum sensing at the beginning of Phase 2. When there is no sensing error, the sensing
outcome for the $m$th ad-hoc band is denoted as $X_m(\alpha T_f)=y_m\in\{0,1\}$ with
$m=1,\ldots,M$. This additional sensing outcome can be utilized in our spectrum sharing design problem to reduce traffic collisions, at the cost of extra computation and signaling delay in Phase 2, as shown in Fig. \ref{fig99}. For notational simplicity, the duration of this extra computation and signaling delay is also assumed to be $\delta T_f$.

Let us define a network state information
(NSI) as
$$\bm\omega\triangleq\{g_n^{s,d},g_n^{s,r},g_n^{r,d},x_m,y_m,~n\in\mathcal {N},m\in\mathcal {M}\},$$ which includes both the channel estimation
and spectrum sensing results over the two phases. In the ergodic setting, the NSI $\bm\omega$ varies
across frames. We assume that the channel fading gains and the ad-hoc
traffic are stationary and ergodic;
furthermore, their statistical distributions are known to the
destination node prior to transmissions.

Similar to \eqref{eq105}, the average traffic collision time for a frame with NSI $\bm\omega$ is determined by
{\small\begin{eqnarray}\label{eq71}
 I(\bm\omega)
\!\!\!\!\!\!\!\!\!\!&&=\sum_{m=1}^M\!\!\left(\!\int_{\bigcup\limits_{n\in\mathcal
{N}_m}\mathbb{I}_n^{(1)}(\bm\omega)} \!\!\!\! \!\!\!\!
\Pr\!\left\{\!X_m(t)\!=\!1|X_m(0)\!=\!x_m\!\right\} \!dt
\right.\nonumber\\&&~~~\left.
+
\int_{\bigcup\limits_{n\in\mathcal {N}_m}\mathbb{I}_n^{(2)}(\bm\omega)}
\!\!\!\!\!\!\!\!\!\!\Pr\!\left\{\!X_m(t)\!=\!1|X_m(\alpha T_f)\!=\!y_m\!\right\}
\!dt\!\right),
\end{eqnarray}$\!\!$}
where $\mathbb{I}_n^{(i)}(\bm\omega)$ denotes the set of CRN
transmission time over sub-channel $n$ in phase $i$ given the NSI
$\bm\omega$. Note from \eqref{eq71} that, in contrast to
\eqref{eq105}, the collision time in Phase
2 now depends on the sensing outcome $y_m$. Since the NSI is stationary and ergodic across
the frames, the long term average traffic collision time can be obtained by taking the expectation of
$I(\bm\omega)$ over the distribution of the NSI
$\bm\omega$ \cite{BK:Resnick}, i.e., {\small
\begin{eqnarray}\label{eq34}
\overline{I}
\!\!\!\!\!\!\!\!&&=\mathbb{E}_{\bm\omega}\!\left[\!\sum_{m=1}^M\!\!\left(\!\int_{\bigcup\limits_{n\in\mathcal
{N}_m}\mathbb{I}_n^{(1)}(\bm\omega)} \!\!\!\! \!\!\!\!
\Pr\!\left\{\!X_m(t)\!=\!1|X_m(0)\!=\!x_m\!\right\} \!dt\!\right.\right.\nonumber\\
&&~~~\left.\left.+\!
\int_{\bigcup\limits_{n\in\mathcal {N}_m}\mathbb{I}_n^{(2)}(\bm\omega)}
\!\!\!\!\!\!\!\!\Pr\!\left\{\!X_m(t)\!=\!1|X_m(\alpha T_f)\!=\!y_m\!\right\}
\!dt\!\right)\!\right].
\end{eqnarray}$\!\!$}

Let us define the transmission time fractions $\theta_n^{(i)}(\bm\omega)$ as in \eqref{eq2}, and define
\begin{eqnarray}
\hat\theta_m^{(i)}(\bm\omega)\triangleq\max\left\{\theta_n^{(i)}(\bm\omega),n\in
\mathcal {N}_m\right\},
\end{eqnarray}
for $m=1,\ldots,M$ and $i=1,2$. It is not difficult to show that the
optimal spectrum access strategies stated in Lemma \ref{lem1} also hold
true for the case with two spectrum sensings in each frame:
\begin{lemma}\label{lem3}
For any given transmission time fractions
$\{\hat\theta_m^{(1)}(\bm\omega)\in[0,\alpha-\delta],\hat\theta_m^{(2)}(\bm\omega)\in[0,1-\alpha-\delta]\}_{m=1}^M$
we have that:
\begin{enumerate}
\item The
optimal spectrum access strategy of Phase 1 is given by
$\mathbb{I}_n^{(1)}(\bm\omega)=[\delta T_f,(\delta+\hat\theta_m^{(1)}(\bm\omega))T_f]$
$(
\mathbb{I}_n^{(1)}(\bm\omega)=[(\alpha-\hat\theta_m^{(1)}(\bm\omega))T_f,
\alpha T_f])$ for all $n\in\mathcal {N}_m$, if the sensing outcome
of Phase 1 is $x_m = 0$ $(x_m = 1)$;
\item The
optimal spectrum access strategy of Phase 2 is given by
$\mathbb{I}_n^{(2)}(\bm\omega)=[(\alpha+\delta)
T_f,(\alpha+\delta+\hat\theta_m^{(2)}(\bm\omega))T_f]$
$(\mathbb{I}_n^{(2)}(\bm\omega)=[(1-\hat\theta_m^{(2)}(\bm\omega))T_f,
T_f])$ for all $n\in\mathcal {N}_m$, if the sensing outcome of Phase
2 is $y_m = 0$ $(y_m = 1)$.
\end{enumerate}
\end{lemma}
An example of Lemma \ref{lem3} is illustrated in Fig. \ref{fig99}.

Define the two functions
\begin{eqnarray}\label{eq78}
\!\!\!\!\!\!\!\!\!\!\!\!\!\!\!&&~~~\hat{\phi}_{(2)}(\theta;y_m=0)\nonumber\\
\!\!\!\!\!\!\!\!\!\!\!\!\!\!\!&&=\!\int_{[(\alpha+\delta)
T_f,(\theta+\alpha+\delta) T_f]} \Pr(X_m(t)\!=\!1|X_m(\alpha T_f)=0)d
t\!\nonumber\\
\!\!\!\!\!\!\!\!\!\!\!\!\!\!\!&&=\!\frac{\lambda}{\lambda\!+\!\mu}\left\{\theta
\!+\!\frac{e^{-({\lambda\!+\!\mu})\delta T_f}}{(\lambda+\mu)T_f}
\!\!\left[\!e^{-({\lambda\!+\!\mu})\theta
T_f}\!-\!1\!\right]\!\!\right\}\!T_f,\\
\!\!\!\!\!\!\!\!\!\!\!\!\!\!\!&&~~~\hat{\phi}_{(2)}(\theta;y_m=1)\nonumber\\
\!\!\!\!\!\!\!\!\!\!\!\!\!\!\!&&=\int_{[T_f-\theta
T_f,T_f]} \Pr(X_m(t)=1|X_m(\alpha T_f)=1)d
t\!\nonumber\\
\!\!\!\!\!\!\!\!\!\!\!\!\!\!\!&&=\!\frac{\lambda}{\lambda\!+\!\mu}\!\left\{\!\theta +\!\frac{\mu}{\lambda}\frac{e^{-({\lambda\!+\!\mu})(1\!-\!\alpha)T_f}}{(\lambda\!+\!\mu)T_f}
\left[e^{({\lambda\!+\!\mu})\theta T_f}\!-\!1\right]\!\!\right\}\!T_f,\label{eq79}
\end{eqnarray}
where $\theta\in[0,1-\alpha]$. In accordance with Lemma \ref{lem3}, $\overline{I}$ in \eqref{eq34} can be
simplified as
\begin{eqnarray} \label{eq36}
\overline{I} \!=\! \mathbb{E}_{\bm\omega}\left\{\!\sum_{m=1}^M\! \left[\!
\phi_{(1)}
\!\left(\!\hat\theta_m^{(1)}(\bm\omega);x_m\!\right)\!+\!\hat{\phi}_{(2)}
\!\left(\!\hat\theta_m^{(2)}(\bm\omega);y_m\!\right)\!\right]\!\!\right\},
\end{eqnarray}
where $\phi_{(1)}(\theta,x_m)$ has been defined in \eqref{eq16} and
\eqref{eq166} for $x_m=0$ and $x_m=1$, respectively.

The achievable average rate of the multi-carrier CRN can be shown to
be
\begin{eqnarray}\label{eq47}
\overline{R}_{CRN}=\min\left\{\overline{R}_{1},\overline{R}_{2}\right\},
\end{eqnarray}
where {\small
\begin{eqnarray}\label{eq72}
\!\!\!\!\!\!\!\!\!\!\!\!&&\overline{R}_{1}\!=\!W\!\!\sum_{m\in\mathcal
{M}}\sum_{n\in\mathcal
{N}_m}\!\!\mathbb{E}_{\bm\omega}\!\!\left[{\hat\theta_m^{(1)}}(\bm\omega)
\log_2\!\!\left(\!1\!+\!\frac{
P_{s,n}^{(1)}(\bm\omega)\max\{g_n^{s,r},g_n^{s,d}\}}{\hat\theta_m^{(1)}(\bm\omega)}\!\right)\right.\!\!\nonumber\\
\!\!\!\!\!\!\!\!\!\!\!\!\!\!&&~~~~~~~\left.+{\hat\theta_m^{(2)}}(\bm\omega)
\log_2\left(1+\frac{
P_{s,n}^{(2)}(\bm\omega)g_n^{s,d}}{\hat\theta_m^{(2)}(\bm\omega)}\right)\right],\!\!\\
\!\!\!\!\!\!\!\!\!\!\!\!\!\!&&\overline{R}_{2}\!=\!W\!\!\sum_{m\in\mathcal
{M}}\sum_{n\in\mathcal {N}_m}\!\!\mathbb{E}_{\bm\omega}\!\!\left[\!
{\hat\theta_m^{(1)}(\bm\omega)} \log_2\!\left(\!1\!+\!\frac{
P_{s,n}^{(1)}(\bm\omega)g_n^{s,d}}{\hat\theta_m^{(1)}(\bm\omega)}\!\right)\right.\nonumber\\
\!\!\!\!\!\!\!\!\!\!\!\!\!\!&&~~~~~~~\left.
+ \hat\theta_m^{(2)}(\bm\omega)
\log_2\left(1+\frac{P_{s,n}^{(2)}(\bm\omega)g_n^{s,d}+
P_{r,n}(\bm\omega)g_n^{r,d}}{\hat\theta_m^{(2)}(\bm\omega)}\right)\right],\!\!\label{eq73}
\end{eqnarray}}
and
$P_{s,n}^{(1)}(\bm\omega),P_{s,n}^{(2)}(\bm\omega),P_{r,n}(\bm\omega)$
are the transmission powers for a given NSI $\bm\omega$. We should point out that the average rate $\overline{R}_{CRN}$ is not the ergodic data rate in the Shannon sense, but one achieved by taking the average over many adaptive channel coding blocks.
The relay node will not transmit but queue up its received data from the source node \cite{Madsen05,ZhangXi_relay_effective_capacity}, until the channel quality and sensing outcome in Phase 2 is favorable.

It follows from \eqref{eq36} and \eqref{eq47} that the ergodic
spectrum access and resource allocation problem is
\begin{subequations} \label{eq35}
\begin{align}
\!\!\!\min_{\substack{P_{s,n}^{(1)}(\bm\omega),P_{s,n}^{(2)}(\bm\omega),\\P_{r,n}(\bm\omega),
\hat\theta_m^{(1)}(\bm\omega),\\\hat\theta_m^{(2)}(\bm\omega)}}
\!\!\!&\mathbb{E}_{\bm\omega}\!\!\left\{\!\sum_{m=1}^M \!\!\left[\! \phi_{(1)}\!
\!\left(\!\hat\theta_m^{(1)}(\bm\omega);x_m\!\right)\!+\!\hat{\phi}_{(2)}\!
\!\left(\!\hat\theta_m^{(2)}(\bm\omega);y_m\!\right)\!\right]\!\!\right\}\\
{\rm s.t.}~~~~~&\overline{R}_{1}\geq \overline{R}_{\min},~~\overline{R}_{2}\geq \overline{R}_{\min}\\
\!\!\!&\mathbb{E}_{\bm\omega}\!\left\{\!\sum_{n = 1}^N\! \left[\!{P}_{s,n}^{(1)}(\bm\omega)\!+\! P_{s,n}^{(2)}(\bm\omega)\!\right]\!\right\}\!\leq\! \overline{P}^s_{\max} \\
\!\!\!&\mathbb{E}_{\bm\omega}\left\{\sum_{n = 1}^N P_{r,n}(\bm\omega)\right\}\leq \overline{P}^r_{\max}\\
\!\!\!& P_{s,n}^{(1)}(\bm\omega),P_{s,n}^{(2)}(\bm\omega),P_{r,n}(\bm\omega)\geq 0,~n=1,\ldots,N\nonumber\\
\!\!\!&\label{eq45}\\
\!\!\!& 0\leq\hat\theta_m^{(1)}(\bm\omega)
\leq\alpha\!-\!\delta, 0\leq\hat\theta_m^{(2)}(\bm\omega)\leq
1\!-\!\alpha\!-\!\delta,\nonumber\\
\!\!\!&~~~~~~~~~~~~~~~~~~~~~~~~~~~~m=1,\ldots,M.\label{eq46}
\end{align}
\end{subequations}
\subsection{Solving Problem \eqref{eq35} in Real-time}\label{sec42}
In the sequel, we provide a real-time method to solve Problem \eqref{eq35} following the idea in Section \ref{sec3B}. Specifically, we solve the following dual optimization problem
\begin{eqnarray} \label{eq41}
\max_{\zeta,\sigma,\varepsilon,\eta\geq0}\left(\min_{\substack{(P_{s,n}^{(1)}(\bm\omega),P_{s,n}^{(2)}(\bm\omega),
P_{r,n}(\bm\omega), \hat\theta_m^{(1)}(\bm\omega),
\hat\theta_m^{(2)}(\bm\omega))\in \overline{\mathcal {V}}}}
\overline{L}\right)
\end{eqnarray}
where $\overline{\mathcal {V}}\triangleq\left\{\left(P_{s,n}^{(1)}
(\bm\omega),P_{s,n}^{(2)}(\bm\omega) , P_{r,n}(\bm\omega),
\hat\theta_m^{(1)}(\bm\omega),
\hat\theta_m^{(2)}(\bm\omega)\right)\right|$ $0\leq
\hat\theta_m^{(1)}(\bm\omega)\leq\alpha-\delta,~0\leq
\hat\theta_m^{(2)}(\bm\omega)\leq 1-\alpha-\delta,P_{s,n}^{(1)}(\bm\omega),$
$\left. P_{s,n}^{(2)}(\bm\omega) ,
P_{r,n}(\bm\omega)\geq 0, n\in\mathcal {N},m\!\in\!\mathcal
{M}\right\}$, and
\begin{eqnarray} \label{eq51}
\overline{L}\!\!\!\!\!\!&&\!\!\!\!=
\frac{1}{T_f}\mathbb{E}_{\bm\omega}\!\left\{\!\sum_{m=1}^M \!\left[\!
\phi_{(1)}
\!\left(\!\hat\theta_m^{(1)}(\bm\omega);x_m\!\right)\!+\hat{\phi}_{(2)}
\!\left(\!\hat\theta_m^{(2)}(\bm\omega);y_m\!\right)\!\right]\!\right\}\nonumber\\
\!\!\!\!\!\!\!\!\!\!\!\!\!\!&&+
\frac{\zeta}{W}(\overline{R}_{\min}\!-\!\overline{R}_1)\!+\!\frac{\sigma}{W}(\overline{R}_{\min}-\overline{R}_2)\!\nonumber\\
\!\!\!\!\!\!\!\!\!\!\!\!\!\!&&+\varepsilon\left\{
\mathbb{E}_{\bm\omega}\left\{\sum_{n=1}^N\left[
P_{s,n}^{(1)}(\bm\omega)+P_{s,n}^{(2)}(\bm\omega)\right]\right\}-\overline{P}_{\max}^s \right\}
\nonumber\\
\!\!\!\!\!\!\!\!\!\!\!\!\!\!&&+\eta\left\{
\mathbb{E}_{\bm\omega}\left[\sum_{n=1}^N P_{r,n}(\bm\omega)\right]-\overline{P}_{\max}^r \right\},
\end{eqnarray} is the partial Lagrangian of \eqref{eq35}.

Let us define $\bm\omega_\ell\triangleq\{g_{n}^{s,r}(\ell),g_{n}^{s,d}(\ell),g_{n}^{r,d}(\ell),x_{m}(\ell),y_{m}(\ell),$ $n\in\mathcal
{N},m\in\mathcal {M}\}$ as the NSI of the $\ell$th frame.
Problem \eqref{eq41} is in general non-causal, because it requires the NSI realizations $\{\bm\omega_\ell\}_{\ell=1}^\infty$ of future frames to solve the inner minimization problem. However, by making use of the statistical distribution of the NSI, we can solve Problem \eqref{eq41} in two steps: First, we optimize the dual variables $\bm\nu=(\zeta,\sigma,\varepsilon,\eta)^T$ \emph{off-line} based on only the statistical distribution of the NSI. Then, the primal solution is updated  \emph{on-line} according to the current NSI $\bm\omega_\ell$. Most computation tasks are accomplished in the off-line dual optimization step, leaving only simple computations for real-time primal solution update, as detailed in the subsequent two subsections.
\subsubsection{Off-line Dual Optimization}
Given a dual variable $\bm \nu = (\zeta,\varphi,\varepsilon,\eta)^T$, the optimal solution to the inner minimization problem of \eqref{eq41} for the NSI realization $\bm\omega$ is provided as follows:

The optimal
$\frac{P_{s,n}^{(1)}(\bm\omega)}{\hat\theta_m^{(1)}(\bm\omega)}$,
$\frac{P_{s,n}^{(2)}(\bm\omega)}{\hat\theta_m^{(2)}(\bm\omega)}$ and
$\frac{P_{r,n}(\bm\omega)}{\hat\theta_m^{(1)}(\bm\omega)}$ of the
inner minimization problem of \eqref{eq41} can be exactly obtained
by \eqref{eq59}-\eqref{eq75}, with $P_{s,n}^{(i)},P_{r,n}$ and
$\hat\theta_m^{(i)}$
replaced by $P_{s,n}^{(i)}(\bm\omega),P_{r,n}(\bm\omega)$ and
$\hat\theta_m^{(i)}(\bm\omega)$, respectively. The optimal
$\hat\theta_m^{(1)}(\bm\omega)$ can be obtained by either
\eqref{eq12} or \eqref{eq38}, depending on the sensing result $x_m$.
By \eqref{eq78} and \eqref{eq79}, the optimal
$\hat\theta_m^{(2)}(\bm\omega)$ is given as follows: If $y_m=0$,
\begin{eqnarray}\label{eq39}
\!\!\!\!\!\!\!\!\!\!\!\!&&\hat\theta_m^{(2)}(\bm\omega)\!=\!
\left[\!-\!\frac{1}{(\lambda\!+\!\mu)T_f}\!\ln\!\left\{1\!-\!\frac{\lambda\!+\!\mu}{\lambda}\!
\!\sum_{n\in \mathcal
{N}_m}\!\left[\zeta f\!\!\left(\!g_n^{s,d}\frac{P_{s,n}^{(2)}(\bm\omega)}{\hat\theta_m^{(2)}(\bm\omega)}\!\right)\!\!\right.\right.\right.\nonumber\\
\!\!\!\!\!\!\!\!\!\!\!\!\!\!&&~~\left.\left.\left.+{\sigma}
f\left(g_n^{s,d}\frac{P_{s,n}^{(2)}(\bm\omega)}{\hat\theta_m^{(2)}(\bm\omega)}+g_n^{r,d}\frac{P_{r,n}(\bm\omega)}{\hat\theta_m^{(2)}(\bm\omega)}\!\right)\right]\right\}-\delta\right]_0^{1-\alpha-\delta}\!\!;
\end{eqnarray}
otherwise, for $y_m=1$,
\begin{eqnarray}\label{eq40}
\!\!\!\!\!\!\!\!\!\!\!\!&&\hat\theta_m^{(2)}(\bm\omega)\!=\!\left[\!1\!-\!\alpha\!+\!\frac{1}{(\lambda\!+\!\mu)T_f}\!\ln\!\left\{\!\frac{\lambda\!+\!\mu}{\mu}
\!\!\sum_{n\in \mathcal
{N}_m}\!\!\left[\zeta f\!\!\left(\!g_n^{s,d}\frac{P_{s,n}^{(2)}(\bm\omega)}{\hat\theta_m^{(2)}(\bm\omega)}\!\right)\right.\right.\right.\!\!\!\!\!\nonumber\\
\!\!\!\!\!\!\!\!\!\!\!\!\!\!&&~~~~\!\left.\left.\left.+{\sigma}f\left(g_n^{s,d}\frac{P_{s,n}^{(2)}(\bm\omega)}{\hat\theta_m^{(2)}(\bm\omega)}
+g_n^{r,d}\frac{P_{r,n}(\bm\omega)}{\hat\theta_m^{(2)}(\bm\omega)}\!\right)\right]\!
-\!\frac{\lambda}{\mu}\right\}\right]_0^{1-\alpha-\delta}\!\!.
\end{eqnarray}
By substituting (\ref{eq12}), (\ref{eq38}), (\ref{eq39}) and (\ref{eq40})
into (\ref{eq59})-(\ref{eq75}), the optimal values of
$P_{s,n}^{(1)}(\bm\omega),P_{s,n}^{(2)}(\bm\omega)$ and $P_{r,n}(\bm\omega)$ can then be obtained.

The optimal dual variable $\bm\nu^\star$ is obtained by a series of the subgradient update in \eqref{eq58}, where the subgradient ${\bm h}(\bm\nu_k)$ is determined by
\begin{eqnarray}\label{eq42}
{\bm h}(\bm\nu_k)\!=\!\left[\!\begin{array}{l}(\overline{R}_{\min}-\overline{R}_{1}^\star)/W\\(\overline{R}_{\min}-\overline{R}_{2}^\star)/W\\
\sum\limits_{n=1}^N
\mathbb{E}_{\bm\omega}\left\{P_{s,n}^{(1)\star}(\bm\omega)+P_{s,n}^{(2)\star}(\bm\omega)\right\}-\overline{P}_{\max}^s\\\sum\limits_{n=1}^N
\mathbb{E}_{\bm\omega}\left\{P_{r,n}^\star(\bm\omega)\right\} -\overline{P}_{\max}^r\end{array}\!\right]\!,\!
\end{eqnarray}
where $P_{s,n}^{(1)\star}(\bm\omega)$,
$P_{s,n}^{(2)\star}(\bm\omega)$ and $P_{r,n}^\star(\bm\omega)$ are
the optimal solution of the inner minimization problem \eqref{eq41}
for given dual variable $\bm\nu_k$ and NSI realization $\bm\omega$, and $\overline{R}_{1}^\star$ and
$\overline{R}_{2}^\star$ are the corresponding rate values in
\eqref{eq72} and \eqref{eq73}, respectively.

In the off-line dual optimization procedure, the true NSI realizations $\{\bm\omega_\ell\}_{\ell=1}^\infty$ are not available. However, the statistical distribution of the NSI $\bm\omega$ is available. We can still compute the expectation terms in \eqref{eq72}, \eqref{eq73} and \eqref{eq42} by means of Monte Carlo simulations. In particular, one may randomly generate a set of NSI realizations $\bm\omega$ following the distribution of the channel fading and sensing outcome. Then, the inner minimization problem of \eqref{eq41} is solved for each artificially generated NSI realization $\bm\omega$. The expectation terms in \eqref{eq72}, \eqref{eq73} and \eqref{eq42} can be obtained by averaging over these NSI realizations. Then, we can use the subgradient method in \eqref{eq58} to derive the optimal dual variable $\bm\nu^\star$ off-line.

\subsubsection{On-line Primal Solution Update}
After $\bm\nu^\star$ is obtained, one can compute the optimal transmission parameters of Problem \eqref{eq35} based on the true NSI $\{\bm\omega_\ell\}_{\ell=1}^\infty$. More specifically, in Frame $\ell$, the CRN needs to solve the inner minimization problem of \eqref{eq41} for the optimal dual variable $\bm\nu=\bm\nu^\star$ and real-time NSI $\bm\omega=\bm\omega_\ell$. The computational complexity of this step is much smaller than that of off-line dual optimization, because each subgradient update involves solving the inner minimization problem of \eqref{eq41} many times for each of the artificially generated NSI realizations.
On the other hand, spectrum sensing and channel estimation are
usually performed at spatially separated nodes, and thus additional information exchanges are required to accomplish the computation task.

In the sequel, we present how to update real-time primal solution such that
the computation and signaling delay $\delta T_f$ can be substantially decreased. In practice, the
BS is able to acquire the channel gain
$\{g_{n}^{s,r}(\ell),g_{n}^{s,d}(\ell),g_{n}^{r,d}(\ell)\}_{n=1}^N$ through channel
prediction before Frame $\ell$ starts, if the wireless channel varies slowly
across the frames \cite{ZhangYan_prediction}. Given the channel gain
$\{g_{n}^{s,r}(\ell),g_{n}^{s,d}(\ell),g_{n}^{r,d}(\ell)\}$, the BS can compute the ratios
$\frac{P_{s,n}^{(1)}(\bm\omega_\ell)}{\hat\theta_m^{(1)}(\bm\omega_\ell)}$,
$\frac{P_{s,n}^{(2)}(\bm\omega_\ell)}{\hat\theta_m^{(2)}(\bm\omega_\ell)}$
and $\frac{P_{r,n}(\bm\omega_\ell)}{\hat\theta_m^{(2)}(\bm\omega_\ell)}$
according to \eqref{eq59}-\eqref{eq75} in Frame $\ell-1$, and then feed the results back to the source and relay nodes before the start time of Frame $\ell$. Once the source and relay nodes obtain the sensing outcomes $x_m(\ell)$ and $y_m(\ell)$, they can compute $\hat\theta_m^{(1)}(\bm\omega_\ell)$ and $\hat\theta_m^{(2)}(\bm\omega_\ell)$ according to the closed-form solutions \eqref{eq12}, \eqref{eq38}, \eqref{eq39}, and \eqref{eq40}. By providing the source and relay nodes with $\frac{P_{s,n}^{(1)}(\bm\omega_\ell)}{\hat\theta_m^{(1)}(\bm\omega_\ell)}$,
$\frac{P_{s,n}^{(2)}(\bm\omega_\ell)}{\hat\theta_m^{(2)}(\bm\omega_\ell)}$
and $\frac{P_{r,n}(\bm\omega_\ell)}{\hat\theta_m^{(2)}(\bm\omega_\ell)}$ ahead of time, the signaling procedure will not cause additional control delay. As $\hat\theta_m^{(1)}(\bm\omega_\ell)$ and $\hat\theta_m^{(2)}(\bm\omega_\ell)$ has closed-form solutions, the time delay $\delta T_f$ for computing $\hat\theta_m^{(1)}(\bm\omega_\ell)$ and $\hat\theta_m^{(2)}(\bm\omega_\ell)$ is quite short. Therefore, the total computation and signaling delay is quite short. In each frame, the destination node needs to send $3N$ parameters to the source and relay nodes.
Note that this amount of information exchanges are equal to that of conventional OFDM CRN without spectrum sharing \cite{Madsen05}, where the BS also needs to feed the ratios $\frac{P_{s,n}^{(1)} (\bm\omega_\ell)}{\hat\theta_m^{(1)}(\bm\omega_\ell)}$,
$\frac{P_{s,n}^{(2)}(\bm\omega_\ell)}{\hat\theta_m^{(2)}(\bm\omega_\ell)}$
and $\frac{P_{r,n}(\bm\omega_\ell)}{\hat\theta_m^{(2)}(\bm\omega_\ell)}$ back to the source and relay nodes.

In practice, the source and relay nodes may not be able to compute $\hat\theta_m^{(1)}(\bm\omega_\ell)$ and $\hat\theta_m^{(2)}(\bm\omega_\ell)$ in \eqref{eq12}, \eqref{eq38}, \eqref{eq39}, and \eqref{eq40}, owing to hardware limitations. In this case, the BS can compute two possible values of $\hat\theta_m^{(1)}(\bm\omega_\ell)$ in \eqref{eq12} and \eqref{eq38} in advance for both of the sensing outcomes $x_m(\ell)=1$ and $x_m(\ell)=0$. Similarly, $\hat\theta_m^{(2)}(\bm\omega_\ell)$ in \eqref{eq39} and \eqref{eq40} can also be computed in advance at the
BS for both $y_m(\ell)=0$ and $y_m(\ell)=1$. The BS sends
$\frac{P_{s,n}^{(1)}(\bm\omega_\ell)}{\hat\theta_m^{(1)}(\bm\omega_\ell)}$,
$\frac{P_{s,n}^{(2)}(\bm\omega_\ell)}{\hat\theta_m^{(2)}(\bm\omega_\ell)}$
and $\frac{P_{r,n}(\bm\omega_\ell)}{\hat\theta_m^{(2)}(\bm\omega_\ell)}$ and
the two possible values of $\hat\theta_m^{(1)}(\bm\omega_\ell)$ and $\hat\theta_m^{(2)}(\bm\omega_\ell)$ to the source and relay nodes. After spectrum sensing, the source and relay nodes simply select the values of $\hat\theta_m^{(1)}(\bm\omega_\ell)$ and $\hat\theta_m^{(2)}(\bm\omega_\ell)$ according to the sensing outcomes $x_m(\ell)$ and $y_m(\ell)$, respectively. This strategy has two benefits: First, all computations are carried out at the BS, and source and relay nodes only need to select appropriate transmission parameters. Second, the source and relay nodes can start transmission right after spectrum sensing, which means the computation and signaling delay $\delta T_f$ is negligible. In each frame, the destination needs to send $3N+2M$ parameters to the source and relay nodes, which is slightly larger than the strategy introduced in last paragraph.

\begin{figure*}[!t]
\begin{center}
{\subfigure[][]{\resizebox{0.45\textwidth}{!}{\includegraphics{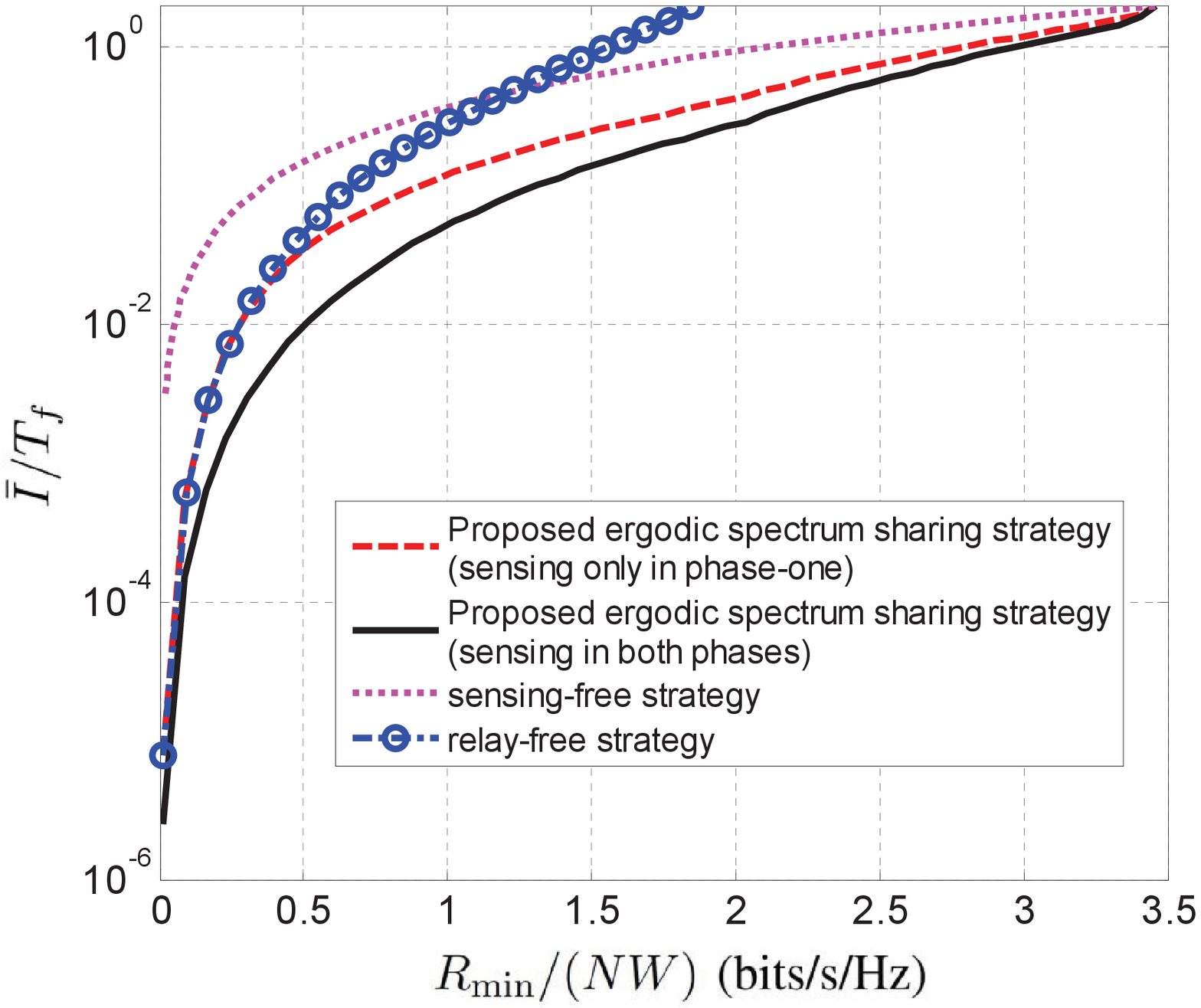}}}}
    \hspace{+0.2cm}
{\subfigure[][ $\bar R_{\rm min}/(NW)=0.6$
bits/s/Hz]{\resizebox{0.45\textwidth}{!}{\includegraphics{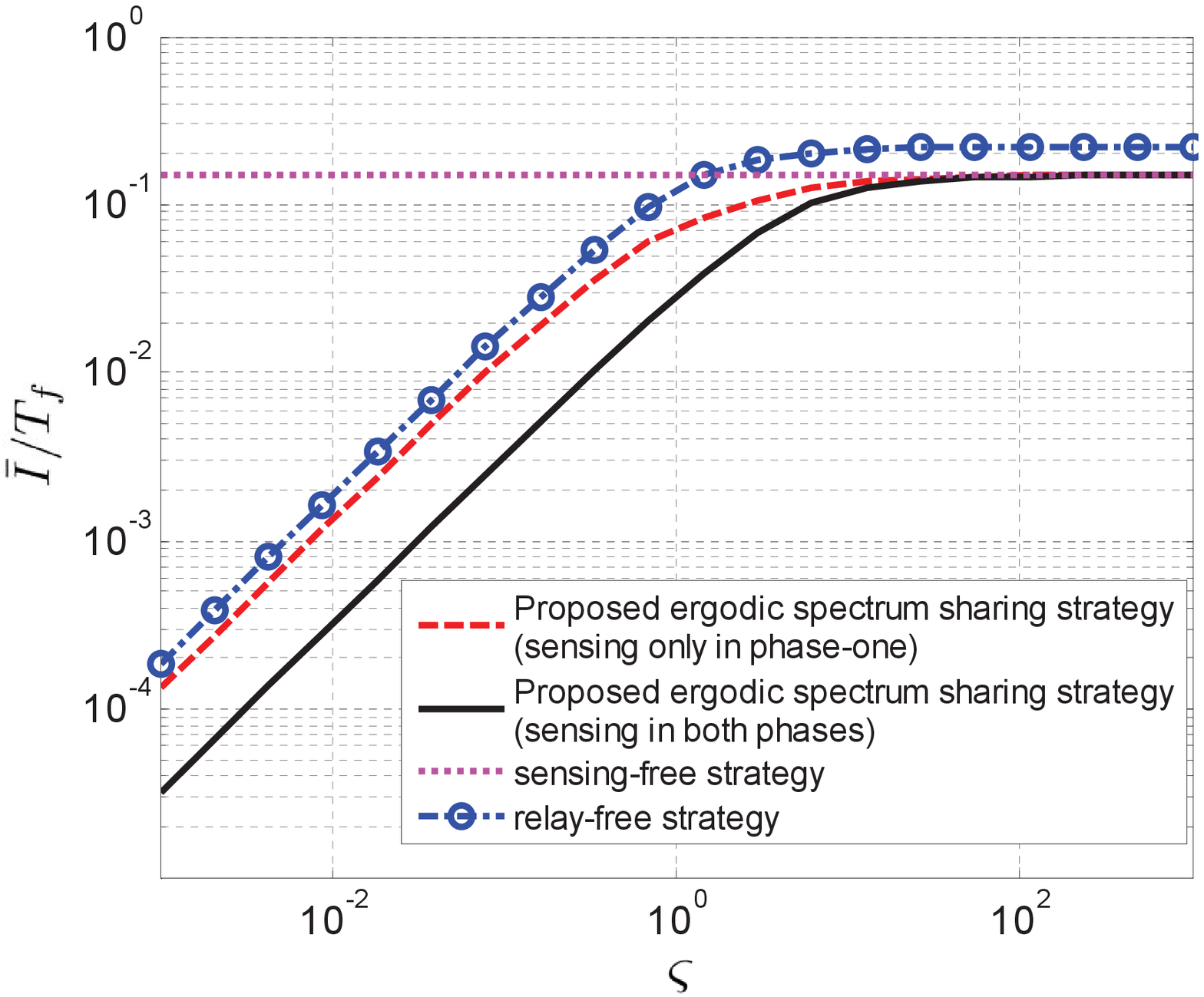}}}
    \vspace{0.5cm}}
{\subfigure[][$\bar R_{\rm min}/(NW)=1.7$
bits/s/Hz]{\resizebox{0.45\textwidth}{!}{\includegraphics{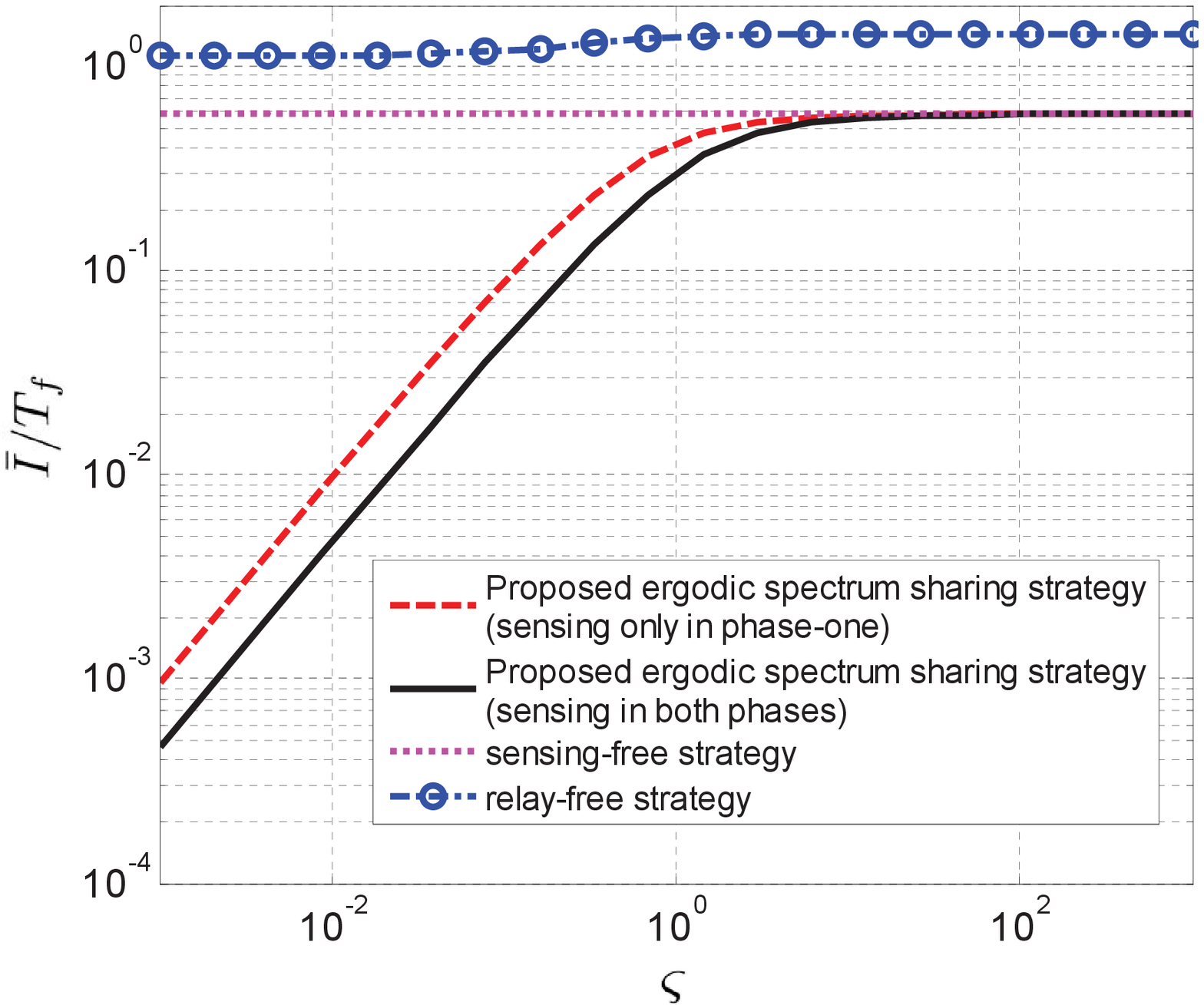}}}}
    \hspace{+0.2cm}
{\subfigure[][$\bar R_{\rm min}/(NW)=2.8$
bits/s/Hz]{\resizebox{0.45\textwidth}{!}{\includegraphics{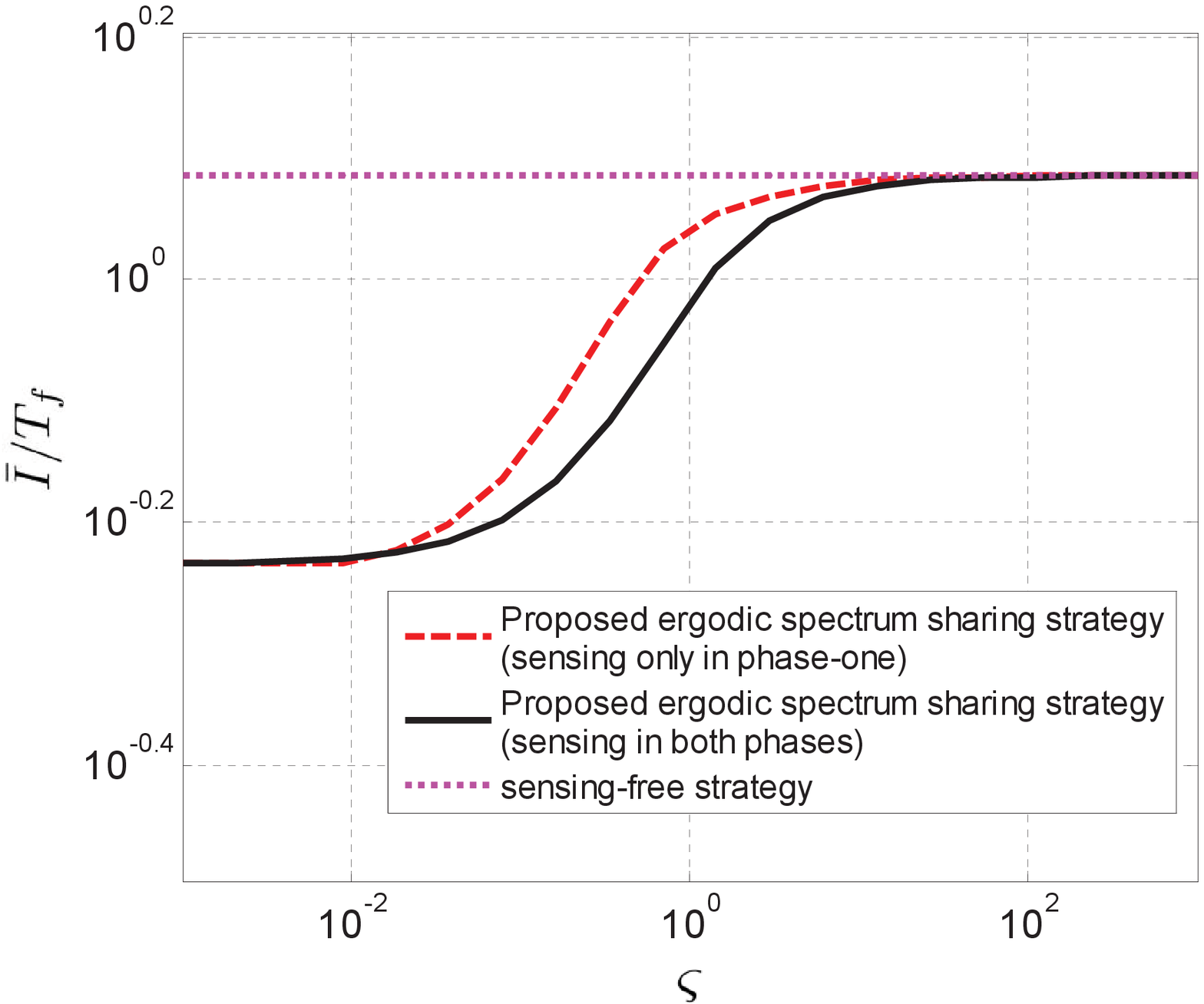}}}}
\vspace{+0.2cm}
\end{center}\vspace{-0.4cm}
\caption{Simulation results of the proposed ergodic transmission control strategy of the CRN. (a) Normalized long term average
collision time ($\bar I/T_f$) versus required long term average uplink
spectrum efficiency $\bar R_{\rm min}/(NW)$, (b)-(d) $\bar I/T_f$ versus
$\varsigma$ for $\lambda=\mu$ and various values of $\bar R_{\rm min}/(NW)$.}\vspace{-0.0cm}
\label{Fig2}
\vspace{-0.7cm}
\end{figure*}
\emph{Remark 2:} With the optimal dual solution $\bm \nu^\star$ obtained ahead of time, the transmission parameters of the CRN in Phase 1 is determined solely by $x_m(\ell)$ but not $y_m(\ell)$. Hence, the primal solution to the ergodic transmission control problem can be computed in a causal manner. However, $y_m(\ell)$ cannot be exploited in the frame-level transmission control problem $(\sf P)$, because neither the dual optimal solution $\bm \nu^\star$ nor the primal optimal solution to $(\sf P)$ can be obtained in advance without $y_m(\ell)$.

\emph{Remark 3:} If the source transmits to the relay node in Phase 1, and the sensing outcome becomes unfavorable in Phase 2, the relay node will queue up its received data from the source node, and wait for better transmission opportunity \cite{Madsen05,ZhangXi_relay_effective_capacity}. In other words, data queuing at the relay node allows the CRN to exploit transmission opportunities in the two phases of each frame separately.

%

\emph{Remark 4:} In practice, sensing error may occur at the source and relay nodes, which means that the ACTIVE (IDLE) ad-hoc traffic state is mistakenly detected as IDLE (ACTIVE). In this case, the source and relay nodes may transmit in different time intervals, leading to extra collisions to the ad-hoc
traffics. In spite of no transmission synchronization in source and relay nodes, the destination node is still able to decode the messages sent from the source and relay nodes. The destination node first decodes the
message from the relay node, and then decodes the source's message, by means of sequential interference cancelation (SIC) decoding \cite{Madsen05}. The performance degradation of the interference metric (i.e., long term average collision time), caused by sensing error, is examined in the next subsection.





\subsection{Simulation Results}\label{sec45}

We present some simulation results in this subsection to demonstrate
the effectiveness of the proposed ergodic CRN spectrum sharing
strategy. The number of sub-channels is $16$ ($N=16$)
and the number of ad-hoc bands is $4$ ($M=4$). Each ad-hoc band
overlaps with 4 consecutive CRN sub-channels, and the 4 ad-hoc
bands do not overlap with each other. The time fraction parameter in each frame $\alpha$ is set to $0.5$.

The channel coefficients $h_n^{i,j}$ (where $i\in\{s,r\}$ and
$j\in\{r,d\},$ $i\neq j$) are modeled as independent and identically
distributed Rayleigh fading with zero mean and unit variance. We assume that the relay node is located right in the middle between the source node and the destination node. The large-scale path loss
factor of all the wireless links is set to 4. Suppose that the
signal-to-interference-plus-noise ratio (SINR) of the
source-destination link is set to
$\frac{P^s_{\max}\mathbb{E}\{|h_n^{s,d}|^2\}}{NWN_n^d}=5$dB. Assuming that $N_n^r=N_n^d$ and $P^s_{\max}=P^r_{\max}$. Then, the SINRs of both the source-relay and relay-destination links are equal to $\frac{P^s_{\max}\mathbb{E}\{|h_n^{s,r}|^2\}}{NWN_n^r}=\frac{P^r_{\max}\mathbb{E}\{|h_n^{r,d}|^2\}}{NWN_n^d}=17$dB according to the path-loss factor.
The simulation results are obtained by averaging over $500$
realizations of NSI (averaging over 500 frames). According to our real-time spectrum sharing strategy in Section \ref{sec42}, the computation and signaling delay is negligible, which means we can choose $\delta = 0$.

\begin{figure}[!t] \centering
    \resizebox{0.5\textwidth}{!}{
    \includegraphics{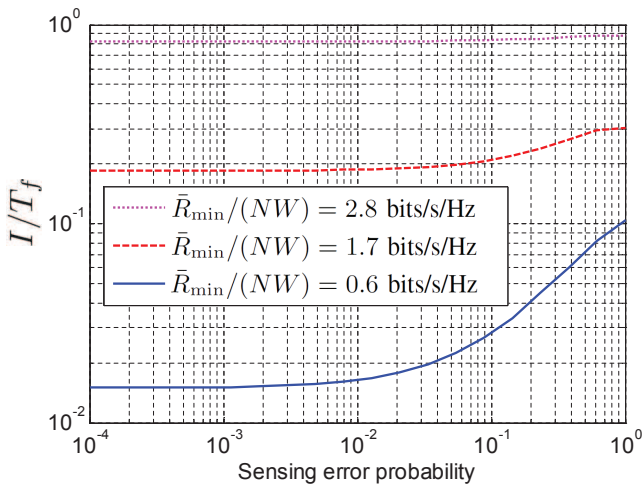}}
    \caption{Simulation results of normalized long term average collision time ($\bar I/T_f$) versus sensing error probability, for different values of $\bar R_{\rm min}/(NW)$.}
    \label{fig6}
    \vspace{0cm}
\end{figure}

Figure \ref{Fig2}(a) shows the simulation results of normalized
long term average collision time ($\bar I/T_f$) versus required long term average uplink
spectrum efficiency $\bar R_{\rm min}/(NW)$, for $\lambda
T_f=\mu T_f=1$ for all $m=1,\ldots,M$. To compare
with the proposed ergodic spectrum sharing strategy, we also perform the same simulations with its relay-free and sensing-free counterparts. The performance of the
ergodic spectrum sharing strategy using only phase-1 spectrum sensing is also
presented. We can observe from Fig. \ref{Fig2}(a) that the proposed
strategy outperforms both the relay-free strategy and the sensing-free strategy.
Moreover, the proposed strategy with spectrum sensing in two phases
performs better than with only phase-1 spectrum sensing.

To further examine how the behavior of the ad-hoc traffic affects
the performance of the proposed strategy, we define a parameter,
called \emph{the relative variation rate of the ad-hoc traffic state} or \emph{the relative sensing period of the CRN}, as
\begin{eqnarray}
\varsigma\triangleq\frac{T_f}{\frac{1}{\lambda}+\frac{1}{\mu}},
\end{eqnarray}
where $\lambda=\lambda$ and $\mu=\mu$ for all $m$. A small value
of $\varsigma$ (that corresponds to small values of $\lambda$ and
$\mu$) implies that the on-off state of the ad-hoc traffic changes slowly in each CRN frame. However, the ad-hoc traffic state would change many times in each CRN frame if
$\varsigma$ is large (that corresponds to large values of $\lambda$
and $\mu$). Figures \ref{Fig2}(b) to \ref{Fig2}(d) show the
simulation results of normalized average collision time ($\bar
I/T_f$) versus relative variation rate $\varsigma$ for $\lambda=\mu$ and various
values of $\bar R_{\rm min}/(NW)$. Since the interference metric of sensing-free strategy is determined by that ratio $\lambda/\mu$, but not how fast the ad-hoc traffic varies (see Eq. \eqref{eq1}), the normalized average collision time of the sensing-free strategy is constant versus $\varsigma$.
From Fig. \ref{Fig2}(b), one can observe that the proposed strategy
performs best. However, the performance gaps between the proposed strategy and the relay-free and sensing-free strategies decrease with $\varsigma$, because the ad-hoc traffic is more difficult to predict for large $\varsigma$. For very large values of $\varsigma$, the proposed strategy has similar performance with the sensing-free strategy. Therefore, spectrum sensing provides no further benefit in this case.

We can also see from Fig. \ref{Fig2}(b) and Fig. \ref{Fig2}(c) that
the performance of the relay-free strategy seriously degrades as $\bar
R_{\rm min}/(NW)$ increases from $0.6$ bits/s/Hz to $1.7$ bits/s/Hz.
The performance degradation of the relay-free strategy is much larger
than that of the proposed strategy because the CRN is capable of
supporting higher uplink throughput than the relay-free strategy. In
Fig. \ref{Fig2}(d), the results of the relay-free strategy are not shown
because this strategy is not feasible in supporting $\bar R_{\rm min}/(NW)=2.8$ bits/s/Hz.

We finally examine the robustness of our proposed strategy against sensing error. Under the same parameter setting associated with the results shown in Fig. 6(a), Fig. \ref{fig6} shows the simulation results of normalized long term average collision time ($\bar I/T_f$) versus sensing error probability, for different values of $\bar R_{\rm min}/(NW)$. One can observe from Fig. \ref{fig6} that, if the sensing error probability is small, e.g., less than $0.01$, the performance degradation of the proposed strategy is insignificant.

\section{Concluding Remarks}\label{sec14}
In this paper, we have investigated optimal spectrum sharing between cooperative relay and ad-hoc networks. Physical-layer resource allocation and MAC-layer spectrum access of the CRN are jointly optimized such that the average traffic collision time
between the two networks is minimized while guaranteeing the CRN throughput requirement. Both frame-level design and ergodic design have been considered. For the latter design, a real-time implementation method of the optimal spectrum sharing strategy has been presented, by exploiting the structure of Lagrangian dual optimization solution. This implementation method has the following merits:
\begin{enumerate}
\item Most computations are accomplished off-line, leaving only simple tasks for real-time computations.
\item Although the sensing outcomes and channel gains are acquired at spatially separate nodes, the information exchange does not cause additional control delay.
\item Almost all the computation loads at the source and relay nodes can be released, at minimal expense of an insignificant amount of information exchanges.
\item Additional spectrum sensing in Phase 2 can be exploited to improve collision prediction. The relay node can queue up its received data if the sensing outcome in Phase 2 is unfavorable, which provides more flexibility for collision mitigation.
\end{enumerate}
Simulation results have been provided to examine the performance of the proposed strategy. We have found that good collision mitigation performance can be achieved if the ad-hoc traffic varies slowly and the required throughput of the relay network is not too high. The presented real-time implementation
techniques may also be useful for real-time transmission control of other wireless networks.

%

\section*{Acknowledgements}
The authors would like to thank P. R. Kumar, Lang Tong, Yongle Wu, Ying Cui and Ness
B. Shroff for constructive discussions about this work.

\appendices
{\setcounter{equation}{0}
\renewcommand{\theequation}{A.\arabic{equation}}
\section{Proof of \emph{Lemma
\ref{lem1}}}\label{Transmission_time_placement}

For $i\in \{1,2\}$, suppose that
$\theta_{n'}^{(i)}=\hat\theta_{m}^{(i)}=\max\{\theta_n^{(i)},
n\in\mathcal {N}_m\}$ for some $n'\in\mathcal
{N}_m$, i.e., sub-channel $n'$ has the longest transmission time
among the sub-channels in $\mathcal{N}_m$. Because
$\mathbb{I}_{n'}^{(i)}\subseteq\bigcup_{n\in\mathcal
{N}_m}\mathbb{I}_{n}^{(i)}$, we have
\begin{eqnarray}\label{eq30}
\!\!\!\!\!\!\!\!\!&&~~\int_{\mathbb{I}_{n'}^{(i)}} \Pr\left\{X_m(t)=1|X_m(0)=x_m\right\}
dt\nonumber\\
\!\!\!\!\!\!\!\!\!&&\leq\int_{\bigcup_{n\in\mathcal {N}_m}\mathbb{I}_{n}^{(i)}}
\Pr\left\{X_m(t)=1|X_m(0)=x_m\right\} dt.
\end{eqnarray}

If the following condition is satisfied
\begin{eqnarray}\label{eq33}
\mathbb{I}_{n}^{(i)}\subseteq \mathbb{I}_{n'}^{(i)},~\forall
n\in\mathcal {N}_m,
\end{eqnarray} for $m=1,\ldots,M$ and $i\in \{1,2\}$, then $\mathbb{I}_{n'}^{(i)} =
\bigcup_{n\in\mathcal {N}_m}\mathbb{I}_{n}^{(i)}$. Then, equality holds in \eqref{eq30}, and the interference is minimized. Therefore, $I$ is determined by only $\mathbb{I}_{n'}^{(i)}$.

We further show that the condition
\begin{equation}\label{eq112}
\mathbb{I}_{n}^{(i)}= \mathbb{I}_{n'}^{(i)},~\forall n\in\mathcal
{N}_m,
\end{equation}
is satisfied at the optimal solution to Problem $(\sf P)$, and $\theta_n^{(i)}=\hat\theta_{m}^{(i)}=\max\{\theta_n^{(i)},
n\in\mathcal {N}_m\}$ for any $n\in\mathcal {N}_m$.

Suppose that \eqref{eq112} does
not hold at the optimal solution to Problem $(\sf P)$. There must exist a sub-channel $k \in \mathcal{N}_m$ such that $\theta_k^{(i)}<\hat \theta_m^{(i)}$. Then, one can
increase $\theta_k^{(i)}$ until $\theta_k^{(i)}=\hat \theta_m^{(i)}$ ($\mathbb{I}_{k}^{(i)}=
\mathbb{I}_{n'}^{(i)}$); this
will increase $R_{CRN}$ in \eqref{eq84} without changing the values of $I$ in
\eqref{eq105}, because $\mathbb{I}_{n'}^{(i)}$ remains the same. In order to reduce $I$, one can further
scale down $\hat \theta_m^{(i)}$ to reduce
$I$ and $R_{CRN}$ simultaneously, until equality holds for the constraint $R_{CRN}\geq R_{\rm min}$. In
summary, if \eqref{eq112} is not true, then one can always achieve a
smaller objective value $I$ for Problem $(\sf P)$. Thus, \eqref{eq112} is satisfied
at the optimal solution to Problem $(\sf P)$, and the remaining problem is to determine the optimal $\mathbb{I}_{n'}^{(i)}$.}

Following the arguments in \cite[Lemma
1]{Geirhofer_Mobile_computing}, one can further show that if
$x_m=0$, the optimal transmission times are given by
$\mathbb{I}_n^{(1)}=[\delta T_f,(\delta + \hat\theta_m^{(1)})T_f]$ and
$\mathbb{I}_n^{(2)}=[\alpha T_f,(\alpha+\hat\theta_m^{(2)})T_f]$
for all $n\in\mathcal {N}_m$, and if $x_m=1$, the optimal
transmission time intervals are given by
$\mathbb{I}_n^{(1)}=[(\alpha-\hat\theta_m^{(1)})T_f, \alpha T_f]$
and $\mathbb{I}_n^{(2)}=[(1-\hat\theta_m^{(2)})T_f, T_f]$ for all
$n\in\mathcal {N}_m$. Lemma \ref{lem1} is thus proved. \hfill
$\blacksquare$

\vspace{-0.2cm} \footnotesize
\bibliography{CR_relay_11}
\begin{IEEEbiography}[{\resizebox{1 in}{!}{\includegraphics{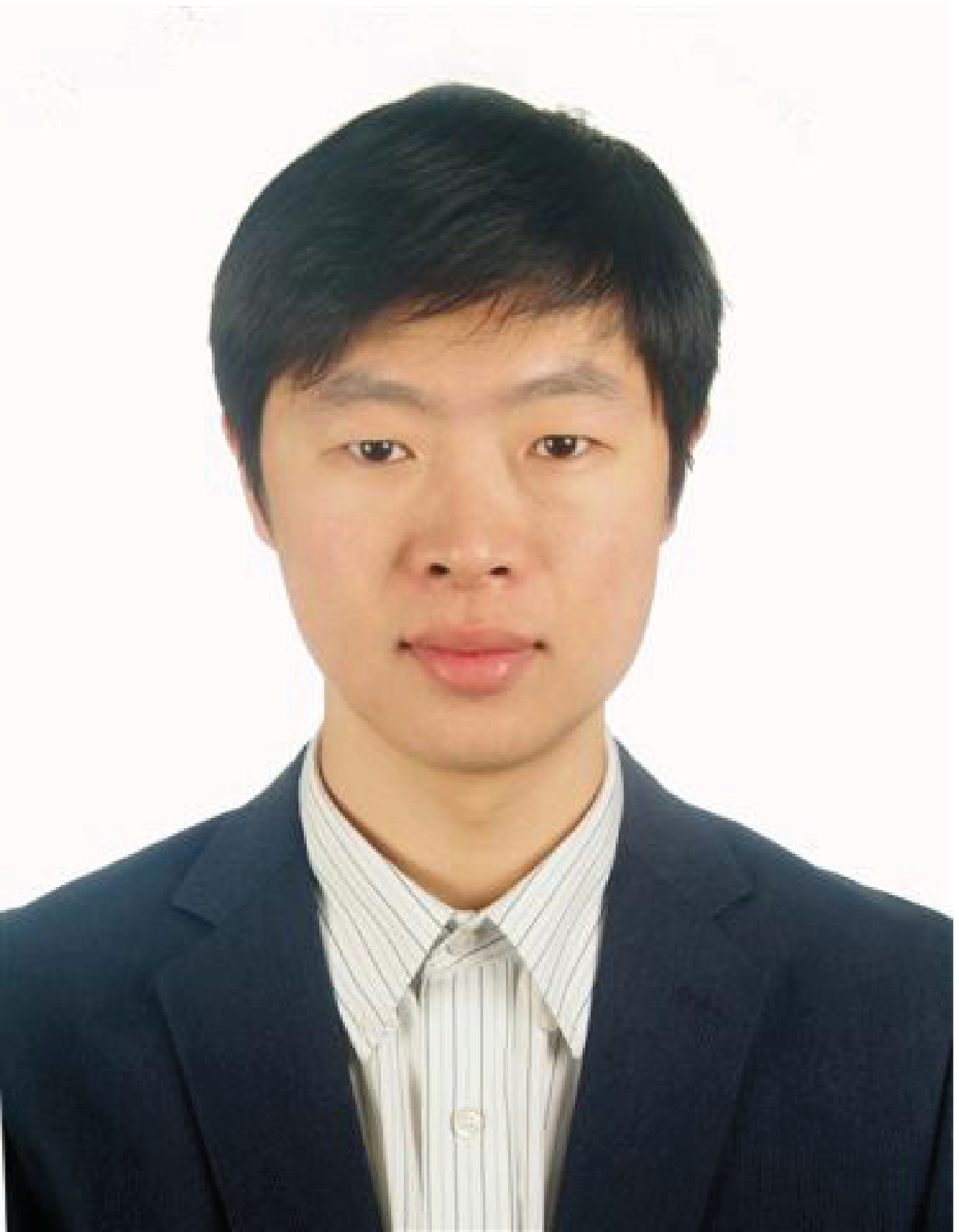}}}]{Yin Sun}
(S'08-M'11) received the B. Eng. degree and Ph.D. degree in electrical engineering from Tsinghua University, Beijing, China, in 2006 and 2011, respectively. He received Tsinghua University Outstanding Doctoral Dissertation Award.
He is currently a Post-doctoral Researcher at the Ohio State University.
His research interests include probability theory, optimization, information theory and wireless communication.
\end{IEEEbiography}

\begin{IEEEbiography}[{\resizebox{1 in}{!}{\includegraphics{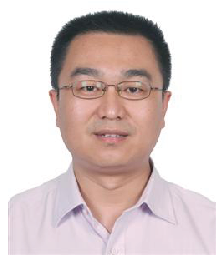}}}]
{Xiaofeng Zhong}(S'02-M'05)
received his Ph.D. degree in Information and Communication System from Tsinghua University in 2005. And he has been the assistant professor in Dept. of Electronic Engineering of Tsinghua University, where he focuses on the MAC and Network protocol design and resource management optimization for wireless ad hoc network, cooperation network and cognitive radio network. Dr. Zhong has published more than 30 papers and own 7 patents.
\end{IEEEbiography}

\begin{IEEEbiography}[{\resizebox{1 in}{!}{\includegraphics{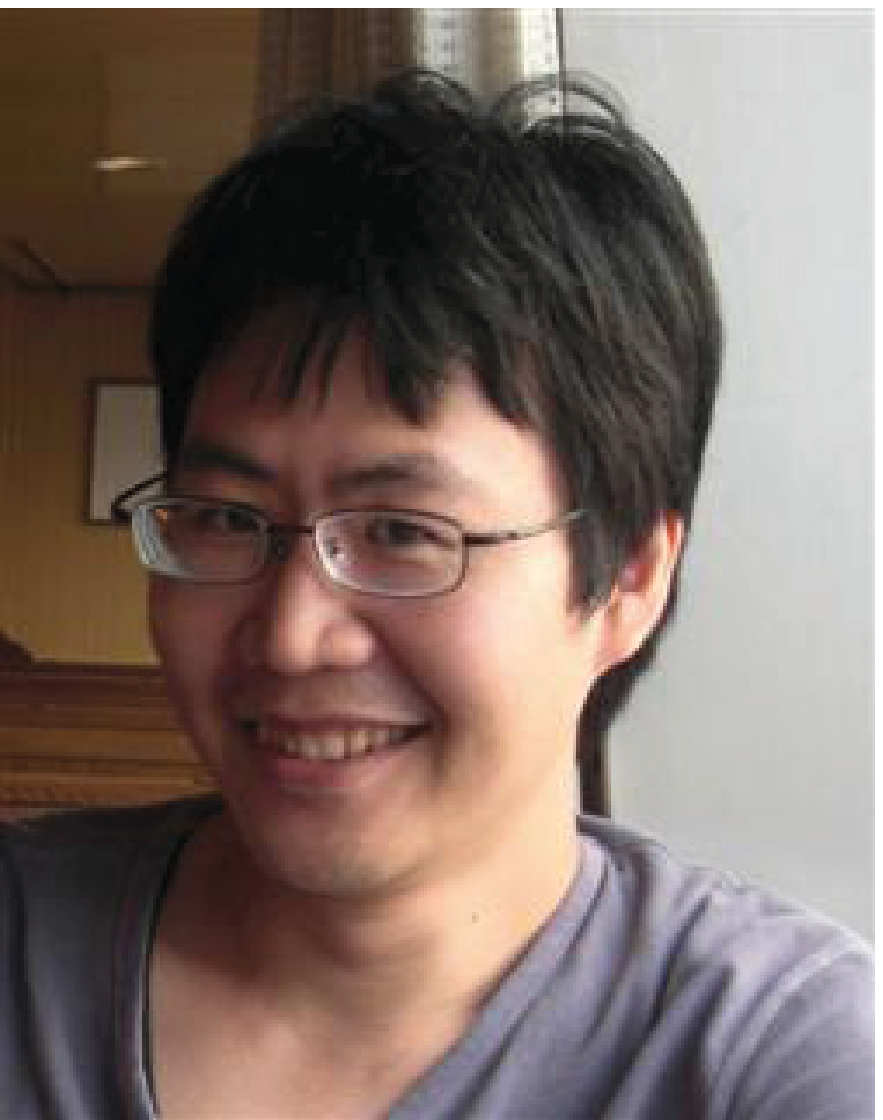}}}]
{\bf Tsung-Hui Chang} (S'07-M'08) received his B.S. degree in
electrical engineering and his Ph.D. degree in communications
engineering from the National Tsing Hua University (NTHU), Hsinchu,
Taiwan, in 2003 and 2008, respectively.
He was an exchange Ph.D. student of University of Minnesota, Minneapolis, MN, USA, a visiting scholar of The Chinese University of Hong Kong, Hong Kong, and a postdoctoral researcher in Institute of Communications Engineering, NTHU. He currently works as a postdoctoral researcher with the Department of Electrical and Computer Engineering, University of California, Davis. His research interests are widely in wireless communications, digital signal processing and convex optimization and its applications.
\end{IEEEbiography}

\begin{IEEEbiography}[{\resizebox{1 in}{!}{\includegraphics{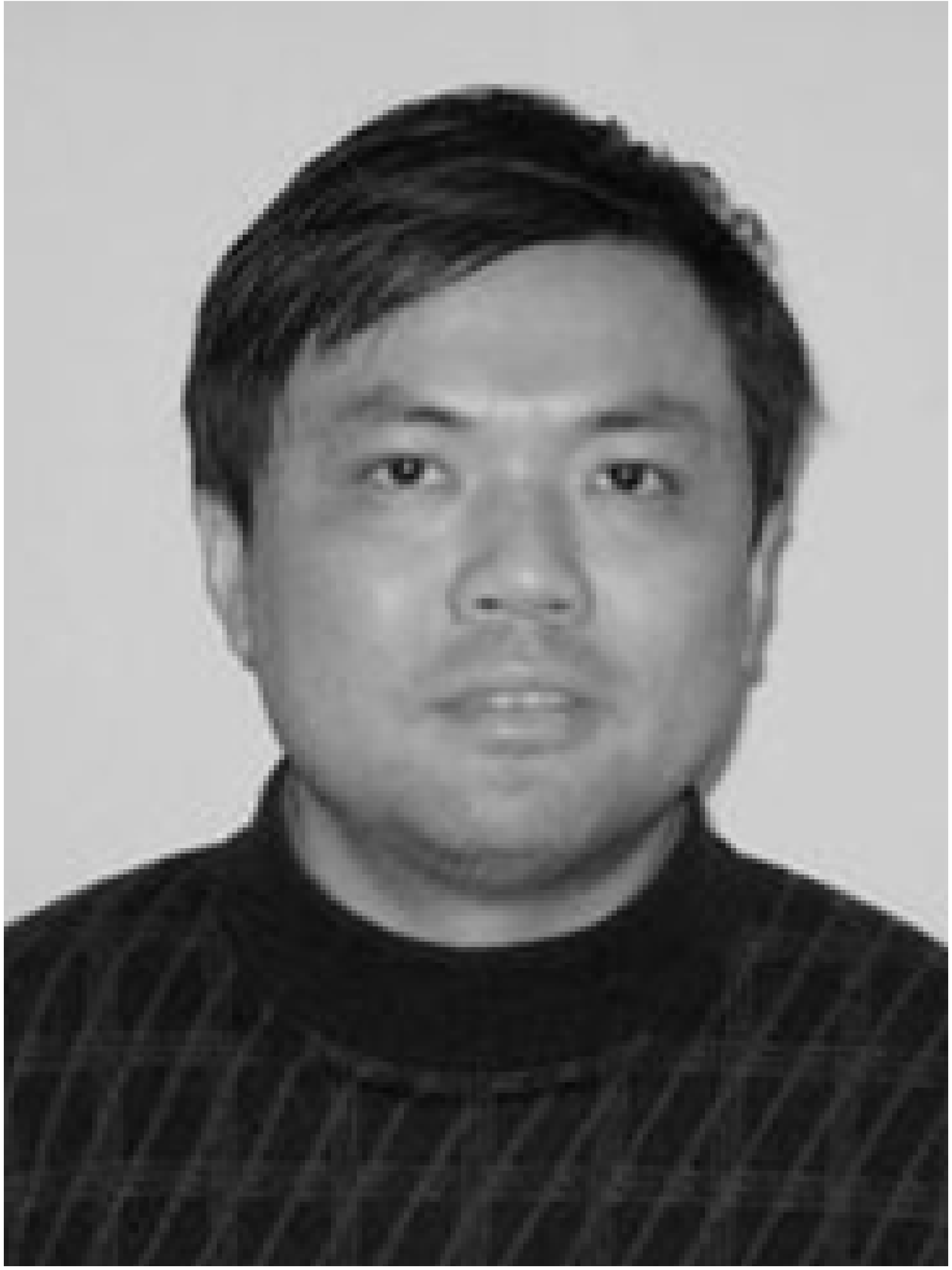}}}] {Shidong Zhou}(M'98)
is a professor at Tsinghua University, China. He received a Ph.D.
degree in communication and information systems from Tsinghua
University in 1998. His B.S. and M.S. degrees in wireless
communications were received from Southeast University, Nanjing, in
1991 and 1994, respectively. From 1999 to 2001 he was in charge of
several projects in China 3G Mobile Communication R\&D Project. He
is now a member of the China FuTURE Project. His research interests
are in the area of wireless and mobile communications.
\end{IEEEbiography}

\begin{IEEEbiography}[{\resizebox{1 in}{!}{\includegraphics{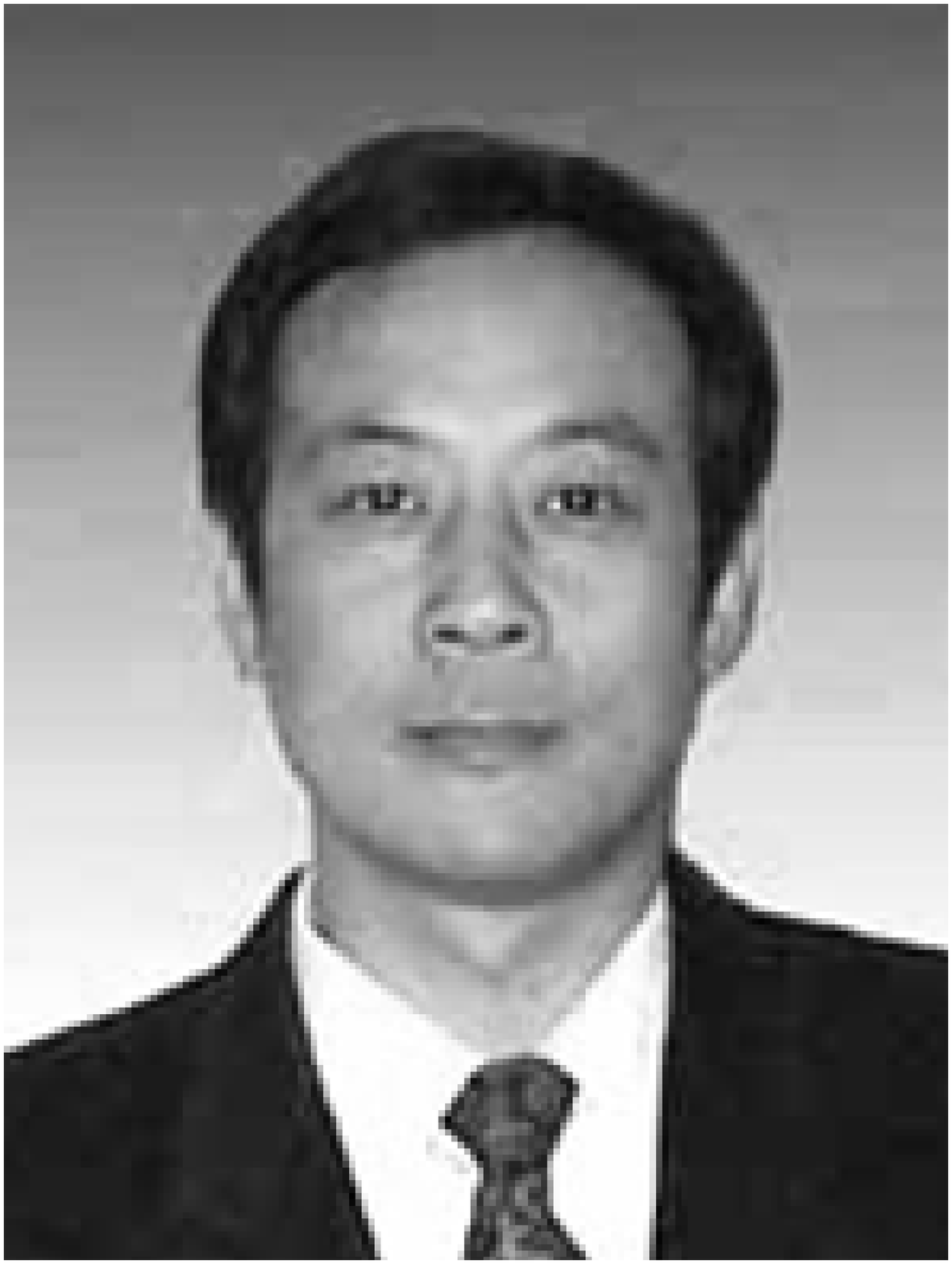}}}]
{Jing Wang}
received B.S. and M.S. degrees in electronic engineering from
Tsinghua University, Beijing, China, in 1983 and 1986, respectively.
He has been on the faculty of Tsinghua University since 1986. He
currently is a professor and the vice dean of the Tsinghua National
Laboratory for Information Science and Technology. His research
interests are in the area of wireless digital communications,
including modulation, channel coding, multi-user detection, and 2D
RAKE receivers. He has published more than 100 conference and
journal papers. He is a member of the Technical Group of China 3G
Mobile Communication R\&D Project. He serves as an expert of
communication technology in the National 863 Program. He is also a
member of the Radio Communication Committee of Chinese Institute of
Communications and a senior member of the Chinese Institute of
Electronics.
\end{IEEEbiography}

\begin{IEEEbiography}[{\resizebox{1 in}{!}{\includegraphics{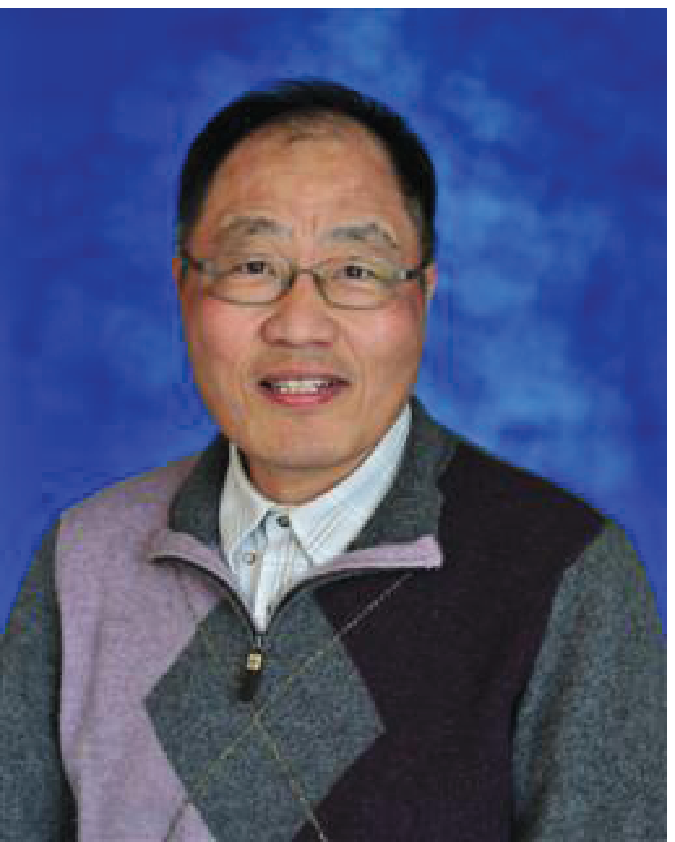}}}]
{Chong-Yung Chi}(S'83-M'83-SM'89)
received the Ph.D. degree in Electrical Engineering from the University of Southern California, Los Angeles, California, in 1983. From 1983 to 1988, he was with the Jet Propulsion Laboratory, Pasadena, California. He has been a Professor with the Department of Electrical Engineering since 1989 and the Institute of Communications Engineering (ICE) since 1999 (also the Chairman of ICE during 2002-2005), National Tsing Hua University, Hsinchu, Taiwan. He has published more than 180 technical papers, including more than 60 journal papers (mostly in IEEE Trans. Signal Processing), 2 book chapters and more than 110 peer-reviewed conference papers, as well as a graduate-level textbook, Blind Equalization and System Identification, Springer-Verlag, 2006. His current research interests include signal processing for wireless communications, convex analysis and optimization for blind source separation, biomedical and hyperspectral image analysis.

Dr. Chi is a senior member of IEEE. He has been a Technical Program Committee member for many IEEE sponsored and co-sponsored workshops, symposiums and conferences on signal processing and wireless communications, including Co-organizer and General Co-chairman of 2001 IEEE Workshop on Signal Processing Advances in Wireless Communications (SPAWC), and Co-Chair of Signal Processing for Communications (SPC) Symposium, ChinaCOM 2008 and Lead Co-Chair of SPC Symposium, ChinaCOM 2009. He was an Associate Editor of IEEE Trans. Signal Processing (5/2001~4/2006), IEEE Trans. Circuits and Systems II (1/2006-12/2007), IEEE Trans. Circuits and Systems I (1/2008-12/2009), Associate Editor of IEEE Signal Processing Letters (6/2006~5/2010), and a member of Editorial Board of EURASIP Signal Processing Journal (6/2005~5/2008), and an editor (7/2003~12/2005) as well as a Guest Editor (2006) of EURASIP Journal on Applied Signal Processing. He was a member of IEEE Signal Processing Committee on Signal Processing Theory and Methods (2005-2010). Currently, he is a member of IEEE Signal Processing Committee on Signal Processing for Communications and Networking.
\end{IEEEbiography}

\end{document}